\documentclass[twocolumn,reprint,superscriptaddress,showpacs,showkeys]{revtex4-1}
\usepackage[utf8]{inputenc}
\usepackage{newtxtext}
\RequirePackage[dvipsnames,usenames]{xcolor}
\usepackage{hyperref}

\usepackage{textcomp}
\usepackage{amsfonts}

\usepackage{amssymb}
\usepackage{amsthm}
\usepackage{bm}
\usepackage{caption}
\usepackage{graphicx}
\usepackage{dcolumn}
\usepackage{bm}
\usepackage{color} 
\usepackage{amsmath,amssymb} 
\usepackage[super]{nth}

\usepackage[english]{babel}

\begin{document}

\title{From Vertices to Vortices in magnetic nanoislands}


\author{Michael Saccone}
\affiliation{Theoretical Division (T4),
Los Alamos National Laboratory, Los Alamos, New Mexico 87545, USA}
\affiliation{Center for Nonlinear Studies (T-Division),
Los Alamos National Laboratory, Los Alamos, New Mexico 87545, USA}

 \author{Jack C. Gartside} 
 \affiliation{Blackett Laboratory, Imperial College London, London SW7 2AZ, United Kingdom, Imperial College London, London SW7 2AZ, United Kingdom}
 \author{Kilian D. Stenning}
 \affiliation{Blackett Laboratory, Imperial College London, London SW7 2AZ, United Kingdom, Imperial College London, London SW7 2AZ, United Kingdom}
\author{ Will R. Branford}
\affiliation{Blackett Laboratory, Imperial College London, London SW7 2AZ, United Kingdom, Imperial College London, London SW7 2AZ, United Kingdom}
\affiliation{
London Centre for Nanotechnology, University College London, London WC1H 0AH, United Kingdom
}

\author{Francesco Caravelli}
\affiliation{Theoretical Division (T4),
Los Alamos National Laboratory, Los Alamos, New Mexico 87545, USA}

\begin{abstract}
Recent studies in magnetic nanolithography show that a variety of complex magnetic states emerge as a function of a single magnetic island's aspect ratio. We propose a model which, in addition to fitting experiments, predicts magnetic states with continuous symmetry at particular aspect ratios and reveals a duality between vortex and vertex states. Our model opens new means of engineering novel types of artificial spin systems, and their application to complex magnetic textures in devices and computing.
\end{abstract}

\maketitle

\section{Introduction}
Remarkable advances in nanofabrication have led to the possibility of engineering designer artificial magnetic systems. In particular, Artificial Spin Ices (ASI) are arrays of magnetically frustrated nanoislands with low-energy state degeneracy which emulate ice models\cite{wang2006artificial,skjaervo2020advances}. Currently, the literature on ASI is expanding from the study of novel phases of matter in many-body physics \cite{farhan2019emergent}, to diverse applications including reconfigurable magnonics \cite{gliga2020dynamics}, neuromorphic \cite{saccone2022direct}
and reservoir computing \cite{gartside2022reconfigurable}.
 The magnetic islands (MIs) of the original Artificial Square Ice \cite{wang2006artificial} had an aspect ratio typically above four, meaning they were four times longer than wide, causing the island to magnetize in a collinear state along the longitudinal direction (a `macrospin' state). Islands with lower aspect ratios can also support the emergence of a vortex domain or even a double-vortex domain, both with zero net magnetization. Understanding the emergence of vortex domains in single islands typically requires the use of costly micromagnetic simulations \cite{gliga2020dynamics}, which numerically integrate the Landau-Lifschitz-Gilbert equation. The emergence of vortices occurs when the aspect ratio is around two to three has been recently used in order to amplify the number of configurations that arrays of nanoislands can have, in  particular granting an advantage for reservoir computing \cite{gartside2022reconfigurable}. Increasingly numerous island states introduce an enriched microstate space and emergent metamaterial memory dynamics\cite{keim2019memory, merrigan2022disorder}. We developed a simple yet feature-rich model to explore the emergence of vortices and guide the ASI field in developing islands that encourage the formation of complex and dynamic magnetic textures.\\
Artificial spin ices can be used to engineer phases of matter since at large aspect ratios, the interaction between fully magnetized (macrospin) nanoislands can be approximated using the dipolar interaction model: collinear islands behave ferromagnetically, while orthogonal ones antiferromagnetically. Since each island in an ASI has two possible magnetic states, one can map a single vertex in a spin ice as an Ising model, with $s_u$, $s_r$, $s_d$ and $s_l$ the Ising states of the up, right, down and left island respectively, with $s_*=\pm 1$.
If we call $J_{\perp}$ the coupling for perpendicular spins,   and $J_{\|}$ for collinear ones, the Hamiltonian of a coordination-four spin ice vertex can be written in the form  $H_{vert}=\frac{1}{2}\Big(J_{\perp}(s_r-s_l)(s_u-s_d)-J_{\|}(s_l s_r+s_d s_u)\Big).$

These interaction patterns can be repeated in any ASI which is a decimation of a square ice. However, as this model fails to accomodate the existence of multiple states in an island, here we argue that a similar but continuous model of a vertex can be used to study the complex features of a single island. \\\
\section{ Model} We employ a generalized bipartite mean field model \cite{contucci} to multiple mean field domains. We partition our system in magnetic domains composed of large numbers of Ising spins. For a magnetic system such a spin ice, one can either choose the magnetic domains to be those of a fully magnetized island, e.g. a macrospin, or those internal to an island. 
 We  assume that our system is partitioned into $K$ subsystems composed of $N$ spins, based on the symmetry of the problem. The first approximation is that each subsystem is composed of perfectly aligned Ising spins, an approximation which is justified for island's aspect ratios above one. The spins in domain $A$ interact with the spins in the domain $B$, with  a certain exchange interaction $J_{AB}$, and among themselves with a certain effective interaction $J_{AA}$.
 The graph which describes the interation between the sets is described by an adjacency matrix ${\mathcal A}_{AB}$, $A=1\cdots K$. Each subset of spins is of size $N_1\cdots N_K$.
 After a brief calculation we can see that, in the spirit of the bipartite case, the Hamiltonian for a larger set of mean field spins is given by
    $H=\sum_{A=1,B=1}^K J_{AB} {\mathcal A}_{AB}\bigg(\sum_{i \in A}^{N_A} s_i\bigg)\bigg(\sum_{j\in B}^{N_B} s_j\bigg)+ \sum_{A=1}^K h_A\bigg(\sum_{i \in A}^{N_A} s_i\bigg)$, where $\mathcal A$ has only zeros and ones, while $J_{AB}$ are coupling constants. As in the case of the bipartite mean field model, we introduce for every set of spins a mean field parameter $m_A=\frac{1}{N} \sum_{i\in A} s_i$, and a Lagrange multiplier $\lambda_A$ to enforce the constraints. Rescaling all the couplings by $J_{AB}\rightarrow J_{AB}/\sqrt{N_{A} N_{B}}$, we then obtained a mean field partition function via Steepest Descent. After a brief calculation, we obtain the emerging mean field equations, given by
$m_A = \tanh(\sum_{B}\beta Q_{AB} m_B+\beta h_A)$ in terms of effective couplings $Q_{AB}=J_{AB} \mathcal A_{AB}$.  

It is interesting at this point to note that the fixed points of these equations are, for sufficiently low temperature, the minima of the Hamiltonian
\begin{eqnarray}
    \mathcal H=\frac{1}{2}\vec m^t \mathbf{ Q} \vec m,
\label{eq:conth}
\end{eqnarray}
but whether or not the minima may be reached depends on the type of dynamics. As we show in the Supp. Mat., this model is able to encode the ground states of various spin ice models.
 One important comment is that if we study spin ice vertices, the effective matrix $\mathbf{ Q}$ has the same structure as the Hamiltonian of a vertex. The key difference is that while the vector $\vec s$ is composed of discrete states, the vector $\vec m$ is composed of continuous magnetic states $-1 \leq m_i\leq 1$. As such, the minimum of the Hamiltonian of eqn. (\ref{eq:conth}) can be studied using its eigenvectors and eigenvalues. When the $m_i$ represent subdomains, the key advantage over micromagnetic simulations is that stable states become a tractable eigenvector problem.

\begin{figure*}[ht!]
    \centering
\includegraphics[width=1\textwidth]{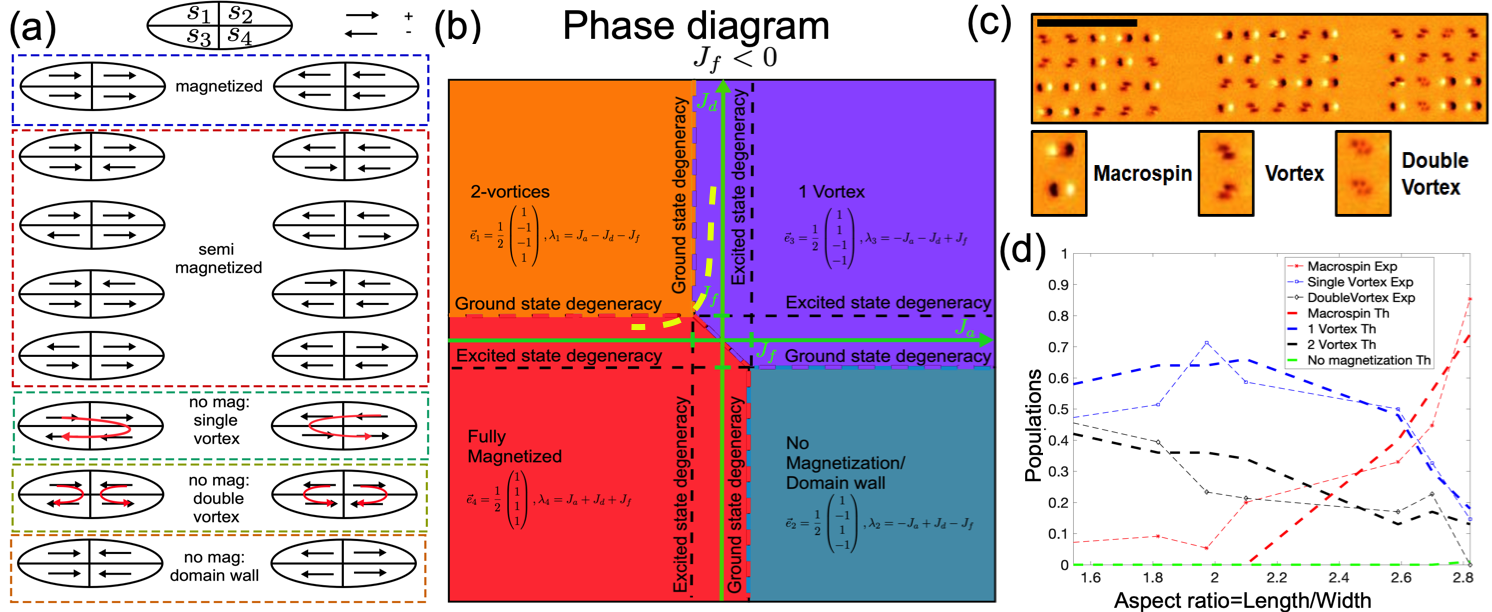}
\caption{\textit{(a)} Magnetic states of a single island and their interpretation. The single vortex and anti-ferromagnetic order are interpreted as real 1 vortex and 2-vortex states. Such nomenclature is consistent with the corresponding relaxed states, with these as initial conditions, obtained from micromagnetic simulations. \textit{(b)} Ground state phase diagram from eqn. (\ref{eq:conth}), with corresponding eigenvectors and eigenvalues. The dashed yellow line represents the behavior of the effective parameters, as a function of the aspect ratio, that fit the experiments. \textit{(c)} MFM image of isolated islands at various aspect ratios, with identified states. The black bar is $5\mu$m. \textit{(d)} Comparison between the experimental populations of island states after AC demagnetization as a function of the aspect ratio, fit by the populations obtained from the mean-field theory, with relaxation given by Metropolis dynamics. }
\label{fig:magisl}
\end{figure*}
\textbf{Magnetic island and eigenvalues of $4\times 4$ structured matrices.}
Let us now restrict our discussion to the behavior of a single MI, which is the building block of all artificial spin ices.
We consider a partitioning of the island as in Fig. \ref{fig:magisl} (a). We assume that the island can be divided in four subdomains $m_1,\cdots,m_4$ assumed to be oriented to the right when positive. By symmetry considerations, we can then introduce three couplings $J_{d}$ between the domain $1$ and $4$ and $3$ and $2$ across the diagonal, $J_f$ for the domains $1$ and $2$ and $3$ and $4$, and then $J_a$ for $1$ and $3$, and $2$ and $4$. Since our effective spins are continuous, one can also introduce a diagonal component $q$. While this interaction is important for the dynamics, since by symmetry considerations $q$ is identical for all the spins, it is not relevant in our considerations below. The interaction matrix is then given by
\begin{eqnarray}\footnotesize
    {\mathbf{Q}}=\left(
\begin{array}{cccc}
 0 & J_f & J_d & J_a \\
 J_f & 0 & J_a & J_d \\
 J_d & J_a & 0 & J_f \\
 J_a & J_d & J_f & 0 \\
\end{array}
\right)
\end{eqnarray}
What is remarkable is that for this matrix, the eigenvectors do not depend on the particular values of $J_a,J_d,J_f$, although clearly the eigenvalues do. The eigenvalues $\eta_i$ and eigenvectors of this matrix are $\eta_1=J_a-J_d-J_f$, $\eta_2=-J_a+J_d-J_f$, $\eta_3=-J_a-J_d+J_f$, $\eta_4=J_a+J_d+J_f$, with corresponding eigenvectors, $\vec v_1=(1, -1, -1,1 )$, 
$\vec v_2= (-1
,1, -1, 1)$
$\vec v_3=(-1, -1,
 1,
 1 )$, $\vec v_4=(1,1, 1, 1)$. Clearly, also the states with opposite signs are valid eigenvectors.\\
  As it turns out, if we look at the graphics of Fig.\ref{fig:magisl} (a), these correspond respectively to: 1) an antiferromagnetic order state, which we interpret as a 2-vortex state; 2) a domain-wall non-magnetized state; 3) a 1-vortex state 4) and a fully magnetized state. To show that this nomenclature is justified, we initialized a micromagnetic simulation in these states, and shown that they relax to the corresponding real vortex states (see Supp. Mat.). If one considers this a squashed square ice vertex, there is a duality between ``vertex type" states and the single island eigenvectors. In fact Type I vertices, 
 correspond to the 2-vortex state. Type II vertices 
 may be constructed from the 1-vortex and fully magnetized states. Type IV vertices, highly energetic magnetic charge states, are non-magnetized state equivalent of the domain wall type. The remaining Type III vertices are superpositions of these eigenvectors, fittingly as they exist as excitations from the spin ice ground state. This simple model incorporates immediately, only via symmetry considerations, the experimentally observed magnetic states of a single MI and those relevant to square ice studies. Interestingly, only four degrees of freedom analytically capture what previously required micromagnetics, despite only crudely approximating the shape of vortices. Finer structures would emerge if more degrees of freedom were allowed through the introduction of a subdivided grid, but these alone present a compelling correspondence to nature. Whether the underlying magnetic interactions are due to exchange, dipolar interactions, or a combination of the two, is irrelevant in this model insofar as the appropriate behaviors are captured. Despite the fact that the entire field of ASI requires this feature, it is often given for granted.\\
Focusing on a single island, what is then important is which eigenvalue is the minimum one as a function of the parameters $J_f,J_a,J_d$. Since we know that one of the states has to be fully magnetized, we can safely set $J_f<0$ and study the minimum eigenvalue problem as a function of $J_a$ and $J_d$ only. The result is shown in the diagram of  Fig.\ref{fig:magisl} (b). In the top right region $J_a>0,J_d>0$, we have as the ground state the 1-vortex state. In the top left region $J_a<J_f$, $J_d>J_f$ we have the two vortex states. The bottom left of the diagram is given by fully magnetized states, while the bottom right by a non-magnetized state, which is however rarely observed. The interface lines between two regions $i,j$ imply that we have eigenvalue degeneracy in the ground state. Thus, on those particular lines $J_a=\pm J_f$, or $J_{d}=\pm J_f$, we have that the magnetic ground state can be written in a \textit{continuous} superposition of the form $\vec v_{ij}(\theta)=\frac{\cos \theta}{2} \vec v_i+\frac{\sin \theta}{2} \vec v_j$ (U(1) symmetry). At the interface between three regions $i,j,k$, the degeneracy becomes triple, and the corresponding eigenvector \textit{at that particular point} can be written in terms of Euler angles as $\vec v_{ijk}(\theta,\phi)=\frac{\cos \theta \cos \phi }{2} \vec v_i+\frac{\sin \theta \cos \phi}{2} \vec v_j+\frac{\sin \phi}{2} \vec v_k$ (O(3) symmetry). The dashed black color lines of Fig.\ref{fig:magisl} (b) are due to the fact that the ground state becomes an excited state, which is then degenerate. The semi-magnetized states shown in Fig. \ref{fig:magisl} (a) are never observed in experiments. One way to explain this phenomenon is that these can only be obtained as a superposition of all the eigenvectors $v_1,\cdots ,v_4$. However, in the ground state phase diagram, there is no point of contact between the four eigenvectors, so these states should only be seen at higher temperatures.\\
\section{Experiments}
Let us now use the model above to intepret the experimental data on islands with various aspect ratio. Permalloy nanomagnets of thickness 25 nm were fabricated via electron beam lithography with $1\mu$m inter-island spacing such that dipolar interaction between neighbouring islands is negligible; in total, we studied 1350 nanoislands with aspect ratios of 0.5-3.0. These were effectively annealed via an AC demagnetisation protocol, and the resulting  states were imaged via magnetic force microscopy (MFM). We then counted the different occurrence of magnetic texture states (macrospin, vortex and double-vortex). A subset of the islands imaged are shown in Fig. \ref{fig:magisl} c). As we can see, as a function of the aspect ratio we only observe states given by a macrospin and single and double vortices \cite{ehrman}. Thus, in the phase diagram of Fig. \ref{fig:magisl} (b) we must be in a region around $J_{a}\approx -J_f$ and $J_{d}\approx J_f$. Because the nanomagnets are athermal at room temperature (the energy barrier between states is much higher than the thermal energy at 300 K), the system can be trapped in a metastable state. To reproduce these results we introduce a form of dynamics for the Hamiltonian of eqn. (\ref{eq:conth}). Note that for $r=1$ the majority of the population is in the 1-vortex state, and for $r\approx2$ we are predominantly in the macrospin state. At intermediate aspect ratios, we observe an increasing population of 2-vortex states. \\
As a result, we hypothesize that parametrically, we are moving along a curve $J_a=-E_a J_f (r+r_a)$, $J_d\propto  E_d \frac{J_f^2}{J_a^2}$. If $E_0=J_f$, then at $J_a=J_f$ we are at the triple point, which allows the coexistence of all states. Such function is shown in Fig. \ref{fig:magisl} (b), which is the yellow dashed curve. A sample of the experimental results is shown  in Fig. \ref{fig:magisl} (c), and the vertex populations for different aspect ratios are shown in Fig. \ref{fig:magisl} (d), averaged over 1350 experiments, with light lines. The theoretically obtained populations after a Metropolis annealing are shown in the same figure (averaged over 100 samples per point), and are given by hard dashed lines. We see a good match between the experimental and the theoretical results, with fitting parameters $E_d=0.6$, $E_a=0.6$, $r_a=1.2$, all of order one.\\
\section{Conclusions}
 We provided a simple model of the emergent states of magnetic nanoislands, as a function of its aspect ratio, showing that these are the eigenvectors of a $4\times 4$ matrix representing the couplings. This seemingly arbitrary coarse graining of a nanoisland not only explained previously seen vortices, but pointed to a new, parametrically robust, two vortex state that we then discovered experimentally. Simultaneously, this justifies the lack of domain wall states and opens the possibility of observing degenerate continuous ground states. By simple symmetry considerations, we attained a mapping between ASI vertex states and MIs states that generated new insight into why magnetic domains form.
As shown in Supp. Mat., our model also reproduces the low energy regime of known spin ices. While micromagnetic simulations are sufficiently fast for understanding properties of 1-10 nanomagnets before an experiment, this mean-field model excels at handling 10-1000 nanomagnets and grasping their non-binary behaviors. Emergent states grant higher memory capacity, analogies to more complex models in statistical physics\cite{schiffer2021artificial}, and the potential to embed a wider class of computational problems within patterned nanomagnet arrays. Additionally, this is the first time a triple-point of macrospin, vortex and double vortex magnetic textures has been observed. Our model also enables dynamical analysis to explain observed emergent phenomena such as avalanches\cite{chern2014avalanches} and coupling to magnon modes\cite{lendinez2019magnetization,urazhdin}.

\textbf{Acknowledgements.}
The work of F.C. and M.S. was carried out under the NNSA of the U.S., DoE at LANL, Contract No. DE-AC52-06NA25396 (LDRD grant - PRD20190195). 
W.R.B. and J.C.G. were supported by the Leverhulme Trust (RPG-2017-257) and the Imperial College London President's Excellence Fund for Frontier Research.
J.C.G. was supported by the Royal Academy of Engineering under the Research Fellowship programme.
We would like to thank C. Nisoli and P. Schiffer for various comments.


\bibliography{bibliography.bib}

\clearpage
\onecolumngrid
\appendix

\section{Micromagnetic Simulation}

\begin{figure*}[ht]
    \includegraphics[width=1\linewidth]{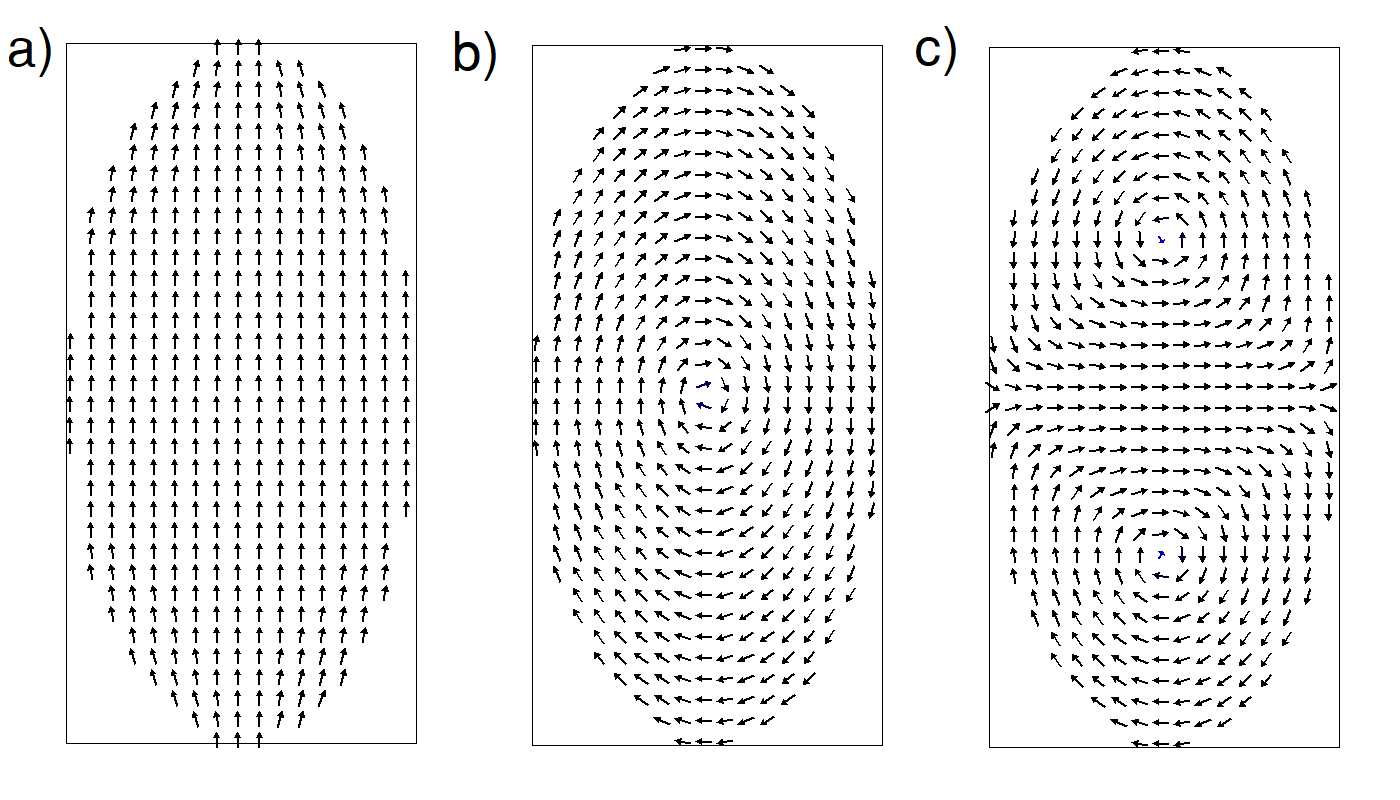}
	\centering
	\caption{Vector fields representing the steady state magnetic moments attained from micromagnetic simulation in OOMMF on a elliptical system of spins with aspect ratio 2. Different seeds of random numbers converge to the a) fully-magnetized, b) 1-vortex, and c) 2-vortex states.
	}
	\label{img:MFDTV}
\end{figure*}

\subsection{Evolution from initial conditions}
\begin{figure}
    \centering
    \includegraphics[scale=0.2]{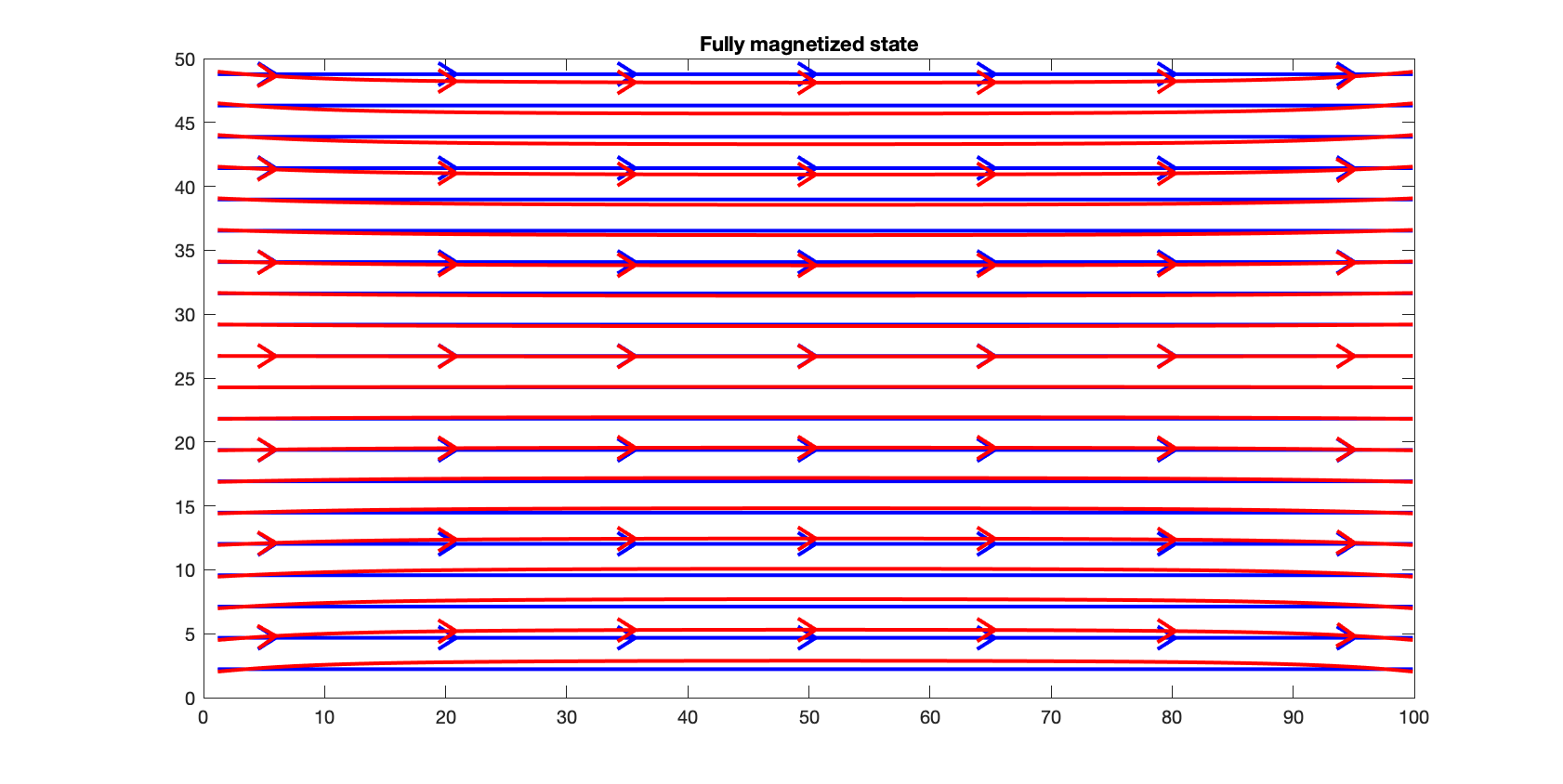}\\
    \includegraphics[scale=0.2]{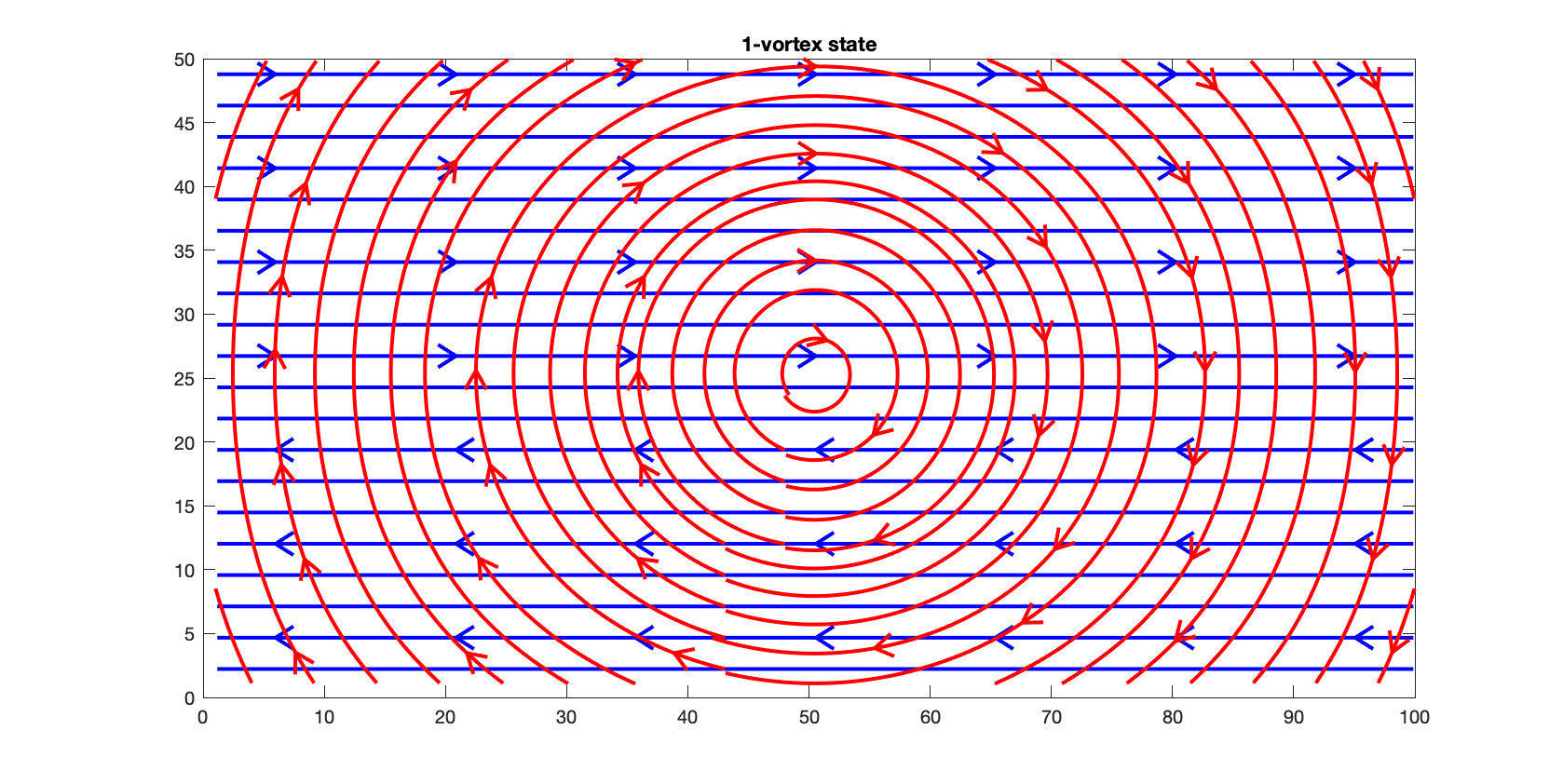}\\\includegraphics[scale=0.2]{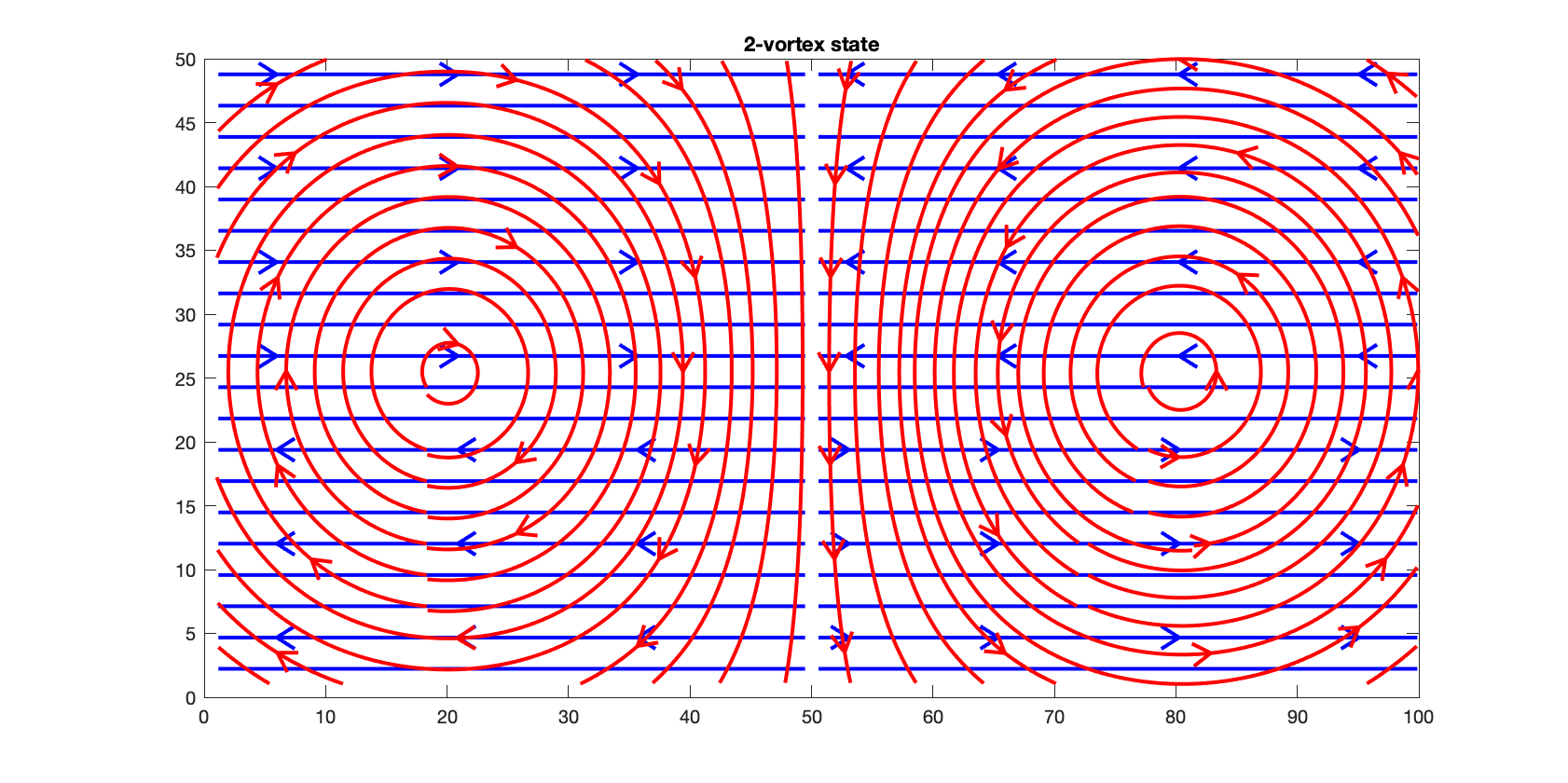}
    \caption{Evolution of the initial states (blue) from our main text into the relaxed magnetic states for a stadium shaped island with aspect ratio 2 (50nm $\times$ 100nm). We see how the system evolves from these states into the fully magnetized, 1-vortex and 2-vortices states respectively. Result obtained with the JuMag.jl package, with the parameters of permalloy.}
    \label{fig:jumag}
\end{figure}

We evolved from the  states of the paper, e.g. respectively the fully-magnetized, 1-vortex and the 2-vortex states, and relaxed a stadium shaped magnetic island. We have used the package JuMag.jl\footnote{https://github.com/ww1g11/JuMag.jl}, written in Julia, in which it is easier to initialize the system. In Fig. \ref{fig:jumag} we can see the initial states (blue lines) versus the relaxed magnetic states. Our simulations were performed on a grid of $100\times 50$ points, with $2nm^2$ cell size, using the parameters for Permalloy. As we can see, these initial states relax into the fully-magnetized, 1-vortex and 2-vortex states respectively (the red lines), showing that our interpretation of the paper is correct from a micromagnetic perspective. We have used random initial conditions for an oval shape matrix of aspect ratio 2, for the parameter of permalloy, using the software OOMMF \footnote{https://math.nist.gov/oommf/}, obtaining similar stable relaxed states. These are shown in Fig. \ref{img:MFDTV}.

\section{Further details on experiments}
Let us now use the model above to intepret the experimental data on islands with various aspect ratio. Permalloy nanomagnets of thickness 25 nm were fabricated via electron beam lithography with $1\mu$m inter-island spacing such that dipolar interaction between neighbouring islands is negligible and islands behave as isolated. 1350 nanoislands were fabricated with lengths of 460-715 nm, widths of 225-275 nm and aspect ratios of 0.5-3.0. The islands were effectively annealed via an AC demagnetisation protocol, initially saturating islands at 65 mT (well-above their coercive fields) and then oscillating down to 0 field in 0.5 mT steps. The resulting island states were imaged via magnetic force microscopy (MFM) and the occurrence of different magnetic texture states (macrospin, vortex and double-vortex) were counted. A subset of the islands imaged are shown in Fig. 1 of the main text.

\clearpage
\section{Derivation of the mean field equations}
The underlying hypothesis for the mean field equations for artificial spin ice is that all nanoislands are composed of Curie-Wei\ss\ domains. We derive mean field quantities representing the average magnetic moment of collections of those domains in equilibrium and equations governing the time evolution of the mean field moments.

\subsubsection{Generalized case: arbitrary graph}
We want to solve a generalization of the Curie-Wei\ss\ model. We have $k=1\cdots K$, and the graph which describes the interation between the sets is described by an adjacency matrix ${\mathcal A}_{AB}$, $A=1\cdots K$. Each subset of spins is of size $N_1\cdots N_K$.

\begin{figure}
    \centering
    \includegraphics[scale=2]{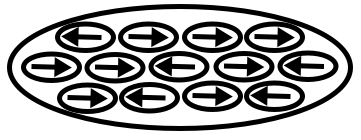}
    \caption{The Wei\ss\  domains in a magnetic nanoisland which we use in our mean field approximation. These are aligned along the longest direction of the island.}
    \label{fig:weiss}
\end{figure}

Here we would like to write down an Hamiltonian which describes this system, with an external field $h_A$ homogeneous on every set, but nonhomogeneous otherwise.

One way to solve it is the following. First, introduce $M_A=\sum_{i\in A} s_i$ and $M_B=\sum_{j\in B} s_j$, and $1$ smartly into the partition function, as
\begin{eqnarray}
    Z&=&\sum_{\{s\}} \int_{-N_A}^{N_A} dM_A\int_{-N_B}^{NB} dM_B\delta(M_A-\sum_{i\in A} s_i)\delta(M_B-\sum_{j\in B} s_j) e^{\beta(J_A (M_A)^2+ J_{AB} M_A M_B + J_B (M_B^2)} \nonumber \\
&=& \sum_{\{s\}}\int_{-N_A}^{N_A} dM_A\int_{-N_B}^{NB} dM_B\int_{-i \infty}^{i\infty} d\lambda \int_{-i \infty}^{i\infty} d\eta  e^{\lambda(M_A-\sum_{i\in A}} s_i)e^{\eta(M_B-\sum_{j\in B} s_j)} e^{\beta(J_A (M_A^2+ J_{AB} M_A M_B + J_B M_B^2)}  \nonumber \\  
&=& \int_{-N_A}^{N_A} dM_A\int_{-N_B}^{NB} dM_B\int_{-i \infty}^{i\infty} d\lambda \int_{-i \infty}^{i\infty} d\eta\  e^{\lambda M_A+\eta M_B+\beta(J_A M_A^2+ J_{AB} M_A M_B + J_B M_B^2)} \nonumber \\
& & \ \ \ \ \ \ \ \ \ \ \ \ \ \ \ \ \ \ \ \ \ \ \ \ \ \ \ \ \ \ \ \ \ \ \ \ \ \ \ \ \ \ \ \ \ \ \ \ \ \ \ \ \ \ \cdot \sum_{\{s_i \in A\}} e^{-\sum_{i\in A} s_i \lambda} \sum_{\{s_j \in B\}} e^{-\sum_{j\in B} s_j \eta}  \nonumber \\  
&=& \int_{-N_A}^{N_A} dM_A\int_{-N_B}^{NB} dM_B\int_{-i \infty}^{i\infty} d\lambda \int_{-i \infty}^{i\infty} d\eta\  e^{\lambda M_A+\eta M_B+\beta(J_A M_A^2+ J_{AB} M_A M_B + J_B M_B^2)}\nonumber \\
& &\ \ \ \ \ \ \ \ \ \ \ \ \ \ \ \ \ \ \ \ \ \ \ \ \ \ \ \ \ \ \ \ \ \ \ \ \ \ \ \ \ \ \ \ \ \ \ \ \ \ \ \ \ \cdot e^{N_A \log 2 \cosh(\lambda)+N_B \log 2 \cosh(\eta)}\nonumber \
\end{eqnarray}
We now rescale $M_A=N_A m_a$ and $M_B=N_B m_B$, and perform the rescaling $J_A=\tilde J_A/2N_A$, $J_B=\tilde J_B/2N_B$, and $J_{AB}=\tilde J_{AB}/\sqrt{N_A N_B}$. We then have
\begin{eqnarray}
    Z&=& \frac{1}{\xi N^2}\int_{-1}^{1} dm_A\int_{-1}^{1} dm_B\int_{-i \infty}^{i\infty} d\lambda \int_{-i \infty}^{i\infty} d\eta\  e^{N \lambda m_A+\xi N \eta m_B+ N\beta(\frac{\tilde J_A}{2} m_A^2+ \frac{\tilde J_{AB}}{\sqrt{\xi}} m_A m_B + \xi \frac{\tilde J_B}{2} m_B^2)}\nonumber \\
& &\ \ \ \ \ \ \ \ \ \ \ \ \ \ \ \ \ \ \ \ \ \ \ \ \ \ \ \ \ \ \ \ \ \ \ \ \ \ \ \ \ \ \ \ \ \ \ \ \ \ \ \ \ \cdot e^{N \log 2 \cosh(\lambda)+\xi N \log 2 \cosh(\eta)}.
\end{eqnarray}
We can now perform the saddle point approximation, writing $Z\approx e^{N f(m_a,m_b,\lambda,\eta)}$, and we obtain the mean field equations:
\begin{eqnarray}
    \partial_{m_A} f=0&\rightarrow&    \lambda+  \beta \tilde J_A m_A+\beta  \frac{\tilde J_{AB}}{\sqrt{\xi}} m_B=0\\
    \partial_{m_B} f=0&\rightarrow&    \eta+  \beta \xi \tilde J_B m_B+\beta  \frac{\tilde J_{AB}}{\sqrt{\xi}} m_A=0 \\
    \partial_{\lambda} f=0&\rightarrow& m_A+\tanh(\lambda)=0 \\
    \partial_{\eta} f=0&\rightarrow& m_B+\tanh(\eta)=0
\end{eqnarray}
which can be rewritten, in terms of $m_A$ and $m_B$, as the equations
\begin{eqnarray}
    m_A&=&\tanh\left[\beta(  \tilde J_A m_A+  \frac{\tilde J_{AB}}{\sqrt{\xi}} m_B)\right]. \nonumber \\
m_B&=&\tanh\left[\beta( \xi \tilde J_B m_B+  \frac{\tilde J_{AB}}{\sqrt{\xi}} m_A)\right].    
\end{eqnarray}
If we define $\theta_A=\beta \tilde J_A$, $\theta_B=\beta \xi \tilde J_B$, and $\theta_{AB}=\frac{\beta \tilde J_{AB}}{\sqrt{\xi}}$, the equations read
\begin{eqnarray}
    m_A&=&\tanh\left(\theta_A m_A+  \theta_{AB} m_B\right). \nonumber \\
m_B&=&\tanh\left( \theta_B  m_B+  \theta_{AB} m_A\right).  
\end{eqnarray}
It is clear from the derivation above that we can extend the same procedure to an arbitrary set of parameters for a Hamiltonian of the form
\begin{equation}
    H=\sum_{A=1,B=1}^K J_{AB} {\mathcal A}_{AB}(\sum_{i \in A} s_i)(\sum_{j\in B} s_j)+ \sum_{A=1}^K h_A(\sum_{i \in A} s_i)
\end{equation}
where $\mathcal A$ has only zeros and ones, while $J_{AB}$ are coupling constants.

As in the case of the bipartite mean field model, we introduce for every set of spins a mean field parameter $m_A=\frac{1}{N} \sum_{i\in A} s_i$, and a Lagrange multiplier $\lambda_A$ to enforce the constraints. Also, we rescale all the couplings by $J_{AB}\rightarrow J_{AB}/\sqrt{N_{A} N_{B}}$. It is not hard to see that the partition function can be written as
\begin{eqnarray}
        Z&=&\int_{-1}^1 [dM] \int_{-i \infty}^\infty [d\lambda] e^{\beta (\sum_{A=1,B=1}^K \sqrt{N_A N_B} J_{AB}{\mathcal A}_{AB} m_A m_B} \nonumber \\
    & &\  \ \ \ \ \ \ \cdot e^{ \sum_{A=1}^K N_A h_A + \sum_{A=1}^K N_A \log 2 \cosh(\lambda_A)}
    \label{eq:effectiveh}
\end{eqnarray}

Now, assuming that $N_A= \xi_A N$ and taking the thermodynamic limit, we get a set of $2K$ mean field equations of the form:
\begin{eqnarray}
    m_A + \tanh(\lambda_A) =0 \nonumber \\
    \lambda_A+\beta \sum_{B} {\mathcal A}_{AB} Q_{AB} m_B+h_A=0
\end{eqnarray}
where $Q_{AB}= \frac{\tilde J_{AB}}{\xi_A \xi_B}$. As a result, we have
\begin{equation}
    m_A = \tanh(\sum_{B} {\mathcal A}_{AB} \beta Q_{AB} m_B+\beta h_A).  
\end{equation}
In the continuous limit of an infinite graph the index $A$ can become a continuous variable. We have thus the result that
\begin{equation}
    m(x)=\tanh\beta\left(\int dy Q(x,y) m(y)+h(x)\right),
\end{equation}
which is a generalization of a single mean field to an infinite set of order parameters. It is not hard to see that it can be obtained from a continuous field theory. Now define the following $ M(y)=\text{atanh}(m(y))$. Thus we have the following nonlinear integral equation of the second kind:
\begin{equation}
    M(x)=\beta h(x)+\beta \int dy Q(x,y) \tanh( M(y)),
\end{equation}
whose solution can be obtained via iterations.

Additionally, we can see that the effective Hamiltonian of eqn. (\ref{eq:effectiveh}) is composed of a mean field energy Hamiltonian given by
\begin{equation}
    H=\frac{1}{2}\sum_{A=1,B=1}^K Q_{AB}  m_Am_B+ \sum_{A=1}^K h_A m_A.
    \label{eq:eff1}
\end{equation}
We will use this Hamiltonian as an effective energy for our islands' interaction.

\section{The case of two islands: bipartite static mean field equations}
The case of two nano-islands features both the Curie-Wei\ss\ temperature of the single islands and the interaction between the islands themselves determining the state of the magnetic moments. Specifically, this is a scenario in which two nanoislands can have both ferro- and antiferro-magnetic interactions, yet behave individually as mean field ferromagnets, as shown in Fig. \ref{fig:weiss}. In the standard treatment of the system with discrete spins: the two islands tend to align head-to-head or head-to-tail at low temperatures. This is essentially right, but as we show below, at the level of mean field, the phase diagram of the two cases is different.

 
 \begin{figure}[b!]
    \centering
    \includegraphics{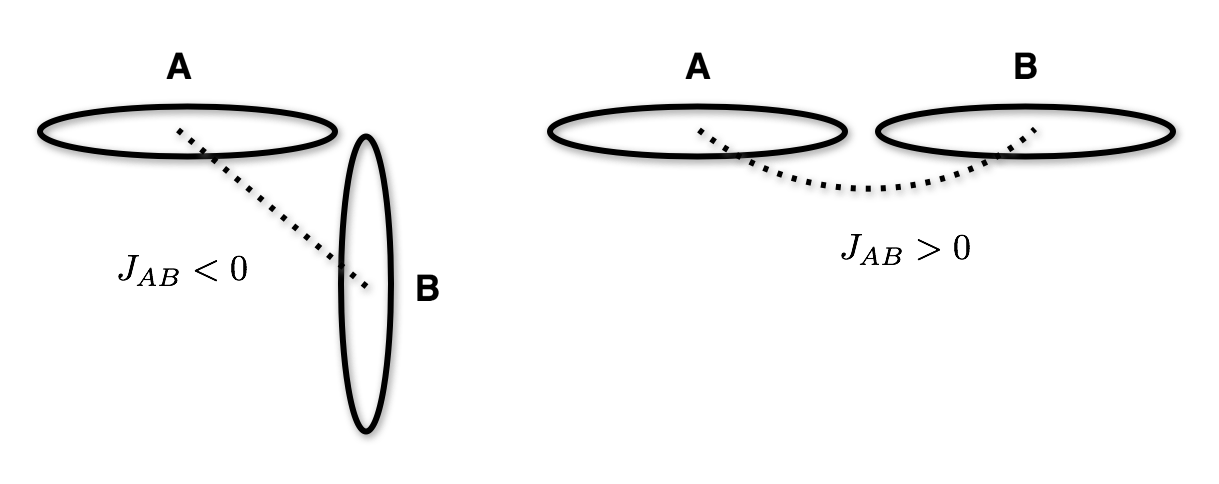}
    \caption{The coupling between the nanoislands: for orthogonal islands it is anti-ferromagnetic, while for parallel it is ferromagnetic.}
    \label{fig:twoislands}
\end{figure}

\begin{figure}
    \centering
    \includegraphics[scale=0.15]{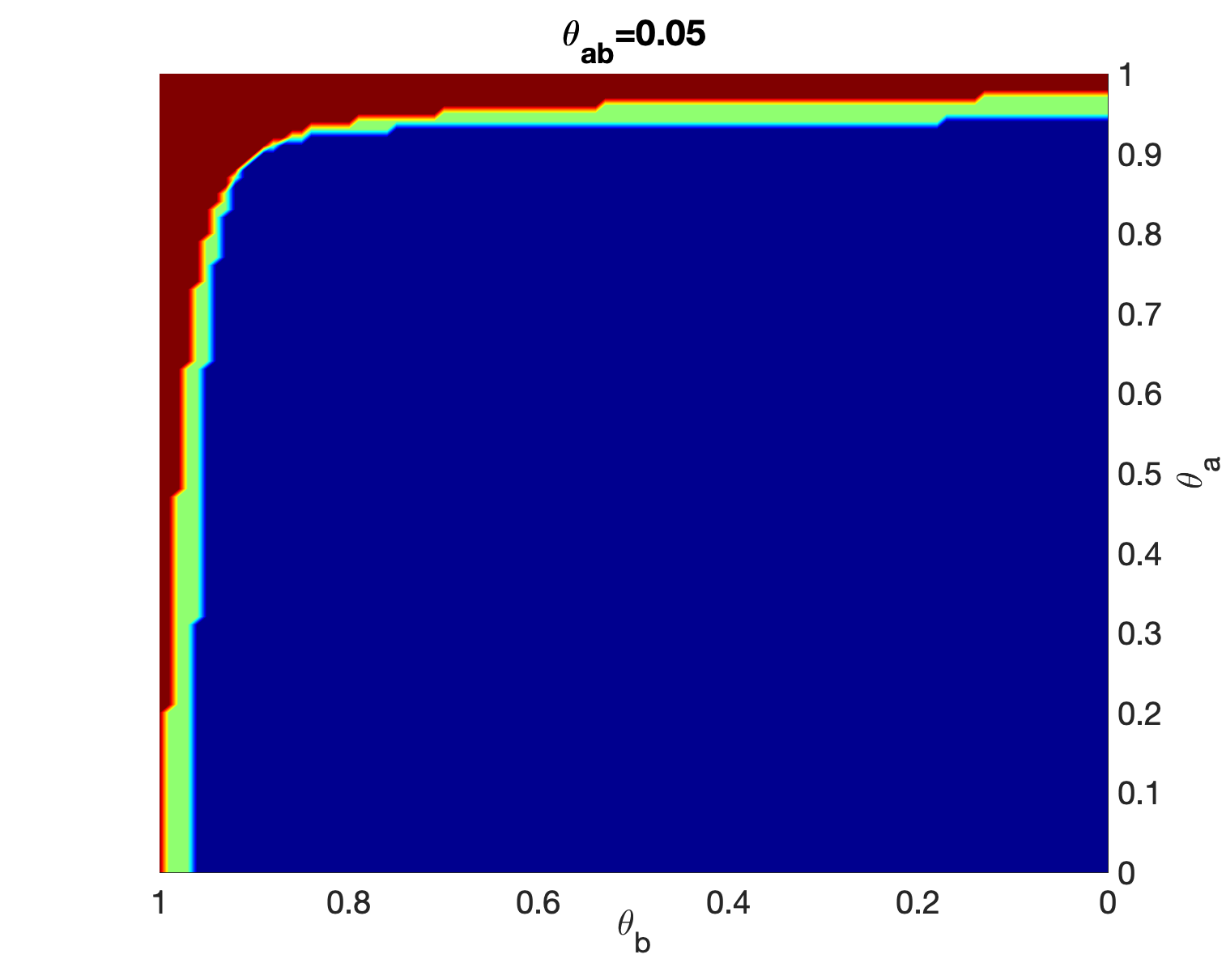}
    \includegraphics[scale=0.15]{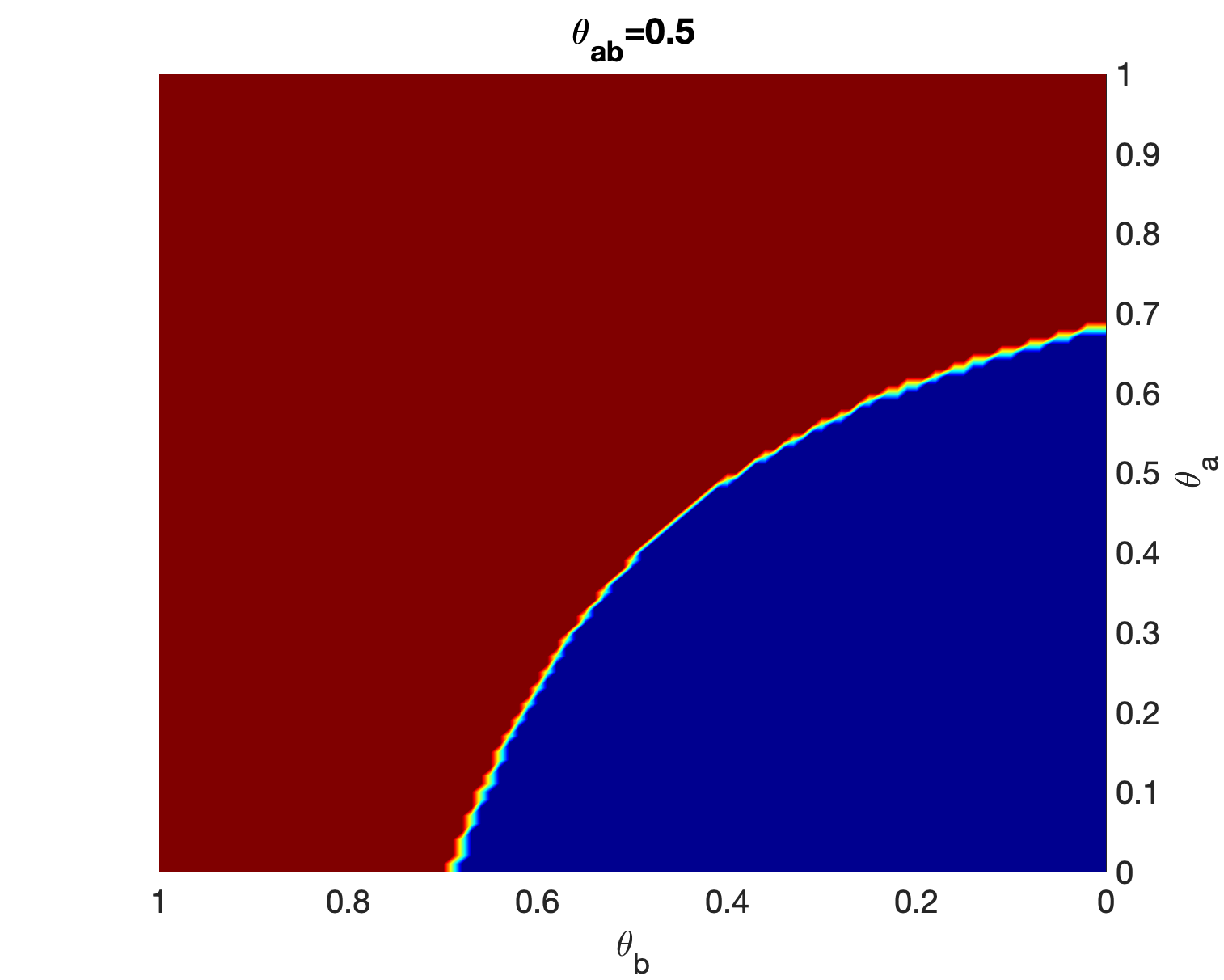}
    \includegraphics[scale=0.15]{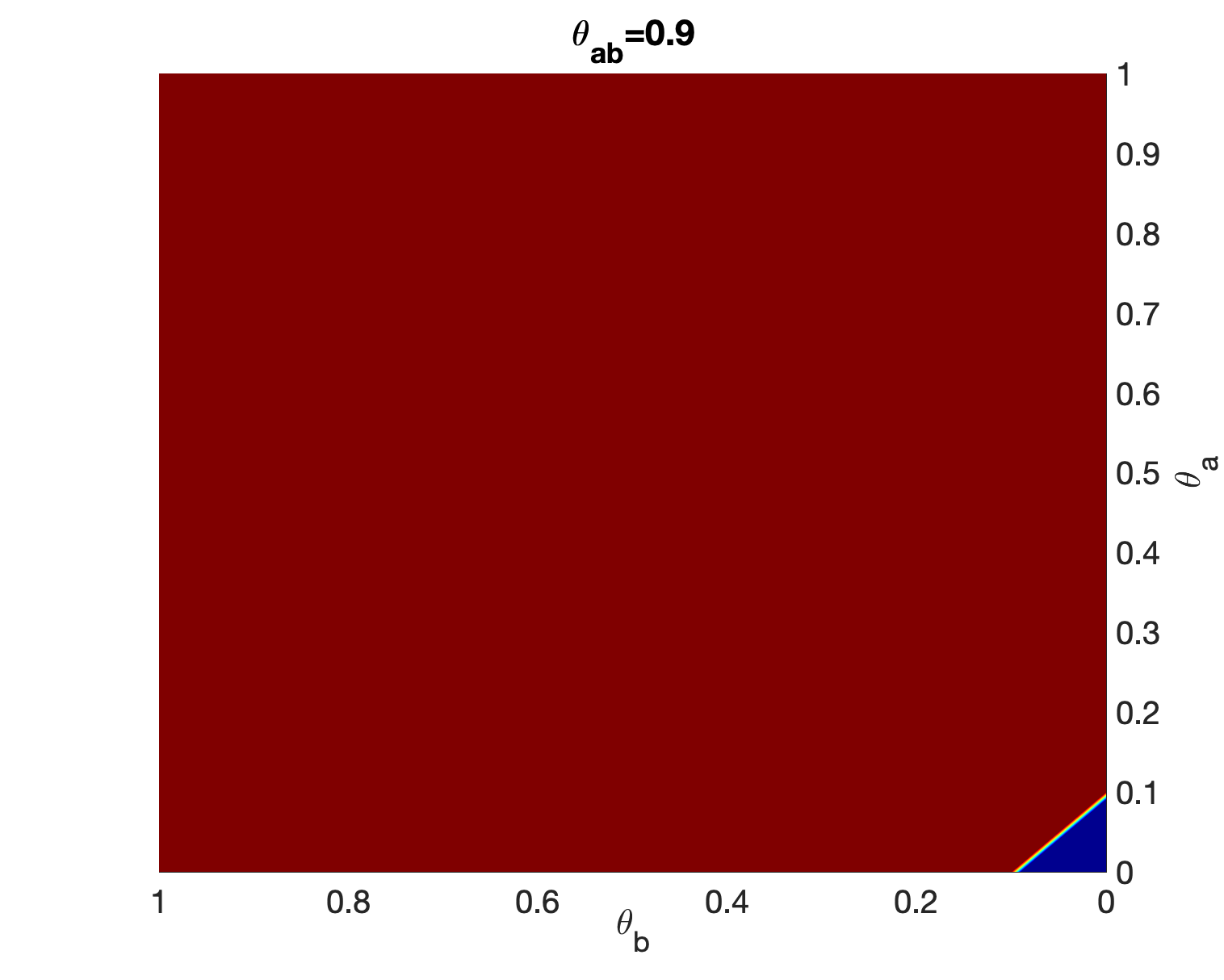}
    \caption{Phase diagram for the bipartite mean field model as a function of $\theta_a$ and $\theta_b$, for $\theta_{ab}\in[0,1]$. Yellow is a phase $3$, green is a phase $2$, while blue is a phase $1$. It is interesting to note that between the phase $1$ and $3$ there is almost everywhere a interfacial phase $2$, aside along the line $\theta_a=\theta_b$, in which there is a direct transition $1$ to $3$. We see that for $\theta_{ab}\rightarrow 1$, the phase $1$ shrinks, leaving space for a phase $3$.}
    \label{fig:dumbell}
\end{figure}

As stated previously,


\begin{eqnarray}
    m_A&=&\tanh\left(\theta_A m_A+  \theta_{AB} m_B\right), \nonumber \\
m_B&=&\tanh\left( \theta_B  m_B+  \theta_{AB} m_A\right),    
\end{eqnarray}
A numerical solution to these equations determines three solutions over different parameter regimes. In phase 1, $m_A = m_B = 0$, the demagnetized state. In phase 2, $m_A = m_B = 1$, the fully ferromagnetic case. 
We can see however that there is a third phase, Phase 3, which can be seen in Fig. \ref{fig:dumbell} (green) at which the system is in an intermediate state $m_A=m_B=1/2$.  While in the top left diagram this is easy to see, this phase shrinks and lies at the boundary between phase $A$ and $B$ for even largers $\theta_{ab}$, and it disappears only when the diagram is in phase 1.

The diagram of what parameters correspond to these phases for $\theta_{ab}=0.05,0.5,1$ is presented in Fig. \ref{fig:dumbell}.

 For $\theta_{AB}<0$ the interfacial region with a phase $2$ disappears, and there is a transition between the phase $1$ and phase $2$ only. See for instance Fig. \ref{fig:dumbell}.

As a result, we see that although the macroscopic notion of head-to-tail and head-to-head behavior of the two islands is essentially correct, at the mean field level one can have interfacial mixed boundaries between the fully magnetized and non-magnetized case as the temperature is lowered, and depending on the interaction strength (distance between the island).

\section{Mean field spin ice: Dynamics}
We consider three models of dynamics for the spin ice effective mean field model: Model A, Metropolis, and Glauber dynamics \cite{hohenberg}. Both in the model A and the Metropolis moves, we enforce $-1\leq m_a\leq1$.
We will compare our results again the Glauber \cite{glauber} dynamics for the discrete spins, with an acceptance probability for a spin flip given
\begin{eqnarray}
    p_{A\rightarrow B}=\frac{e^{-\frac{\Delta E}{\kappa T}}}{1+e^{-\frac{\Delta E}{\kappa T}}}
\end{eqnarray}
with $\Delta E=2 \sum_{b}J_{ab} s_b$.
\subsection{Model A dynamics}
The first model we consider is the so called Model A relaxational dynamics, defined by \cite{hohenberg}
\begin{eqnarray}
    \frac{\partial m_A}{\partial_t}=-\gamma \frac{\delta H}{\delta m_A}+\xi_A(t),
\end{eqnarray}
with $\xi_A(t)$ being a stochastic term with the properties
\begin{eqnarray}
\langle \xi_A(t)\rangle&=&0 \\
\langle \xi_A(t)\xi_B(t^\prime) \rangle&=&2T \gamma \delta_{AB} \delta(t-t^\prime).
\label{eq:diff}
\end{eqnarray}
above, $\gamma$ is the relaxation rate.


The solution to such a linear stochastic differential equation is well known:
\begin{eqnarray}
    \vec s(t)=e^{-\gamma \bm{Q} t} \vec s_0+e^{\gamma \bm{Q} t}\int_0^t ds\ e^{\gamma \bm{Q} s} \vec \xi(s),
\end{eqnarray}
and thus
\begin{eqnarray}
    \langle \vec s(t)\rangle&=&e^{-\gamma \bm{Q} t} \vec s_0+e^{\gamma \bm{Q} t}\int_0^t ds\ e^{\gamma \bm{Q} s} \langle \vec \xi(s)\rangle ,\\
    &=&e^{-\gamma \bm{Q} t} \vec s_0.
\end{eqnarray}
If we consider that $-1\leq s_i(t)\leq 1$, the asymptotic state is the one determined by the maximum \textit{positive} eigenvalue. If the maximum eigenvalue is negative, the asymptotic state is unmagnetized. 

In the case of a four island vertex, these dynamics agree with simple energetic considerations. Considering four islands where $s_i = 1$ when they point up and to the right,

\begin{equation}
    \mathbf{Q} = \begin{pmatrix}
q & -J_1 & J_2 & J_1\\
-J_1 & q & J_1 & J_2\\
J_2 & J_1 & q & -J_1\\
J_1 & J_2 & -J_1 & q\\
\end{pmatrix}
\end{equation}

The eigenvalues and corresponding eigenvectors of this matrix are
\begin{equation}
    \mathbf{v}_1 = \begin{pmatrix}
    1\\
    -1\\
    -1\\
    1
    \end{pmatrix}, 
    \mathbf{v}_2 = \begin{pmatrix}
    1\\
    0\\
    1\\
    0
    \end{pmatrix}, 
    \mathbf{v}_3 = \begin{pmatrix}
    0\\
    1\\
    0\\
    1
    \end{pmatrix}, 
    \mathbf{v}_4 = \begin{pmatrix}
    -1\\
    -1\\
    1\\
    1
    \end{pmatrix}.
\end{equation}
$$\lambda_1 = q + 2J_1 -J_2,$$
$$\lambda_2 = \lambda_3 = q + J_2,$$  
\begin{equation}
    \lambda_4 = q - 2J_1 -J_2.
\end{equation}
If we only consider positive nonzero parameters, the system asymptotically approaches $\pm \mathbf{v}_1$ when $J_1 > J_2$ and a superposition of $\pm \mathbf{v}_2$ and $\pm \mathbf{v}_3$ when $J_2 > J_1$. The former case corresponds to both possible Type I vertices while the fully saturated superpositions of the latter are the four Type II vertices, precisely the local low energy states of square ASI. When $J_1 = J_2$, the eigenvectors abruptly change:
\begin{equation}
    \mathbf{v}_1 = \begin{pmatrix}
    1\\
    -1\\
    0\\
    0
    \end{pmatrix}, 
    \mathbf{v}_2 = \begin{pmatrix}
    1\\
    0\\
    1\\
    0
    \end{pmatrix}, 
    \mathbf{v}_3 = \begin{pmatrix}
    0\\
    1\\
    0\\
    1
    \end{pmatrix}, 
    \mathbf{v}_4 = \begin{pmatrix}
    -1\\
    -1\\
    1\\
    1
    \end{pmatrix}.
\end{equation}
$$\lambda_1 = \lambda_2 = \lambda_3 = q + J_1,$$
\begin{equation}
    \lambda_4 = q - 2J_1 -J_2.
\end{equation}
For all positive and nonzero parameters, all six ice rule obeying vertices are possible as average asymptotes. 

\begin{figure}
    \centering
    \includegraphics[scale=0.35]{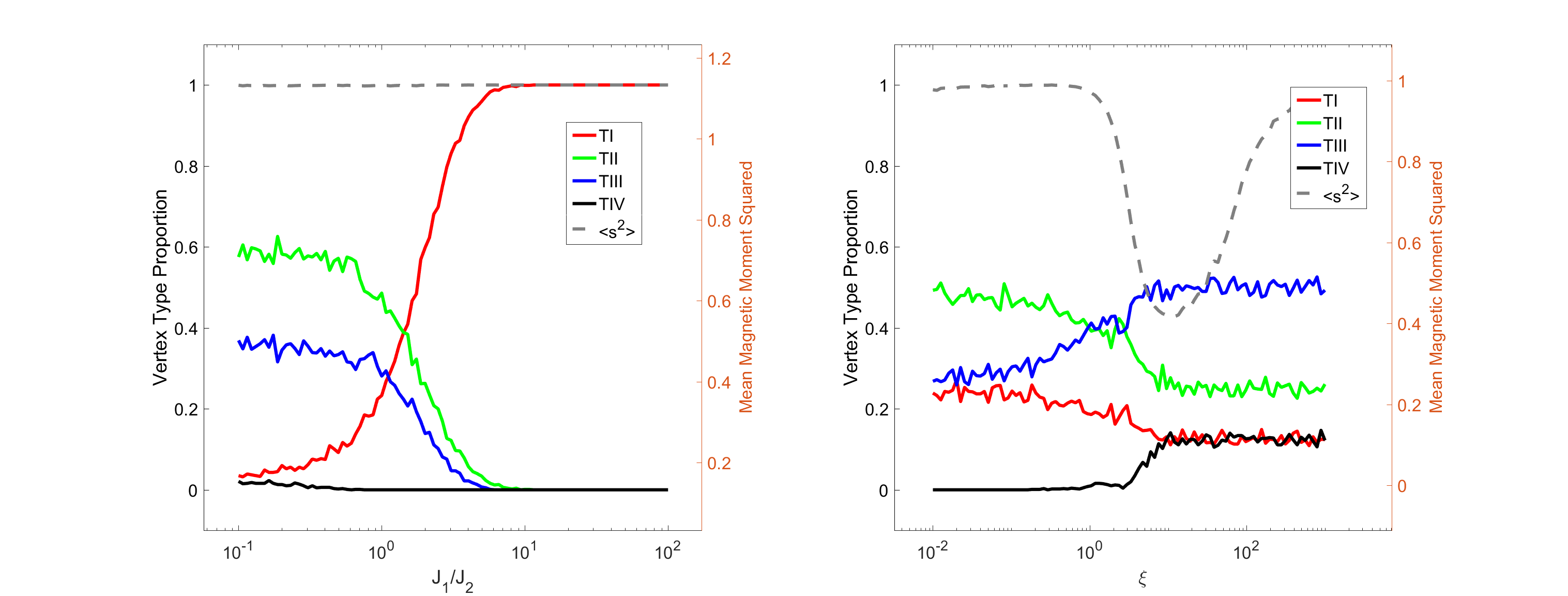}
    \caption{Asymptotic vertex types for model A dynamics. In both plots, $q = 10$ and $J_2 = 1$. Left: $J_1$ is varied while $\xi = .1$. Right: $\xi$ is varied while $J_1 = 1$. }
    \label{fig:modelA}
\end{figure}

To confirm that this is relevant without noise averaging, the stochastic differential equations were Euler integrated and the asymptotic states were recorded in terms of vertex types (Fig.~\ref{fig:modelA}) . $J_1$ was set to a range of parameters in one case, while the square root of the variance of the noise $\xi$ was modified in another case. In all scenarios, $q = 10$ and $J_2 = 1$. When $J_1$ varied, $\xi = .1$, and when $\xi$ varied, $J_1 = 1$. In the former case, Type 1 vertices dominate for high $J_1$, Type 2 dominate and coexist with Type 3 vertices when $J_1$ is low. In the latter case, Type I and III occupy 25\% of asymptotic states and Type II are 50 \% of the states. This gradually transitions to the ergodic arrangements of states at high $\xi$, with an intermediate region where the average spin squared dips to around 50\%. 

\subsection{Metropolis moves}
The Metropolis dynamics, employed in the paper to reproduce the populations of the single vertex, is given a two step move. The first, is a proposed move given by a diffusion of the mean field value, as in eqn.(\ref{eq:diff}). We pick randomly a variable $m_A$,  and propose a diffusive move
\begin{eqnarray}
    m_A\rightarrow m_A^\prime=m_A+\xi
\end{eqnarray}
with $\langle \xi\rangle=0$, and $\langle \xi(t)^2 \rangle=2\kappa T \gamma$. Then, the accept the move with probability given by the Metropolis acceptance rate, 
\begin{eqnarray}
    r=\text{min}(1,e^{-\frac{H-H^\prime}{\kappa T}}),
\end{eqnarray}
which enforces detailed balance. This is the model we simulated in the main text, with $T=J/2$ and $\sigma=0.05$.

\subsection{Mean field Glauber dynamics}
Given the Hamiltonian
\begin{equation}
    H=\sum_{AB} \mathcal A_{AB} J_{AB} \sum_{a\in A} s_a \sum_{b\in B} s_b + \sum_{A} h_A(\sum_{i \in A} s_i)
\end{equation}
In the following we will assume $J_{AB}\equiv J$. We consider the following Glauber dynamics \cite{glauber} for each individual spin. We have 
\begin{eqnarray}
   \frac{ w_a(\vec s)}{w_a(s_a)}&=&\frac{P_{eq}(s_a)}{P_{eq}(s)}=e^{-2 \beta J s_a \sum_b A_{ab} s_b -2\beta h_as_a} \nonumber \\
   &=&\frac{1- s_a \tanh(\beta J \sum_{b\in N(a)} A_{ab}s_b + \beta h_as_a)}{1+ s_a \tanh(\beta J \sum_{b\in N(a)} A_{ab}s_b + \beta h_as_a)}.
\end{eqnarray}
As it is known, a rate which satisfies the detail balance is given by
\begin{equation}
    w_i(s)=\frac{1}{2}\left( 1-s_i \tanh(\beta \sum_{j\in N(i)} A_{ij}s_j + \beta h_is_i) \right)
\end{equation}
The dynamics for each spin is given by
\begin{equation}
    s_i(t+\Delta t)=\begin{cases}
    s_i(t) & p_+=1- w(s) \Delta t\\
    -s_i(t) & p_-=w(s) \Delta t\\
    \end{cases}
\end{equation}
Thus,
\begin{equation}
    \frac{ds_i}{dt}=-2 \langle s_i w(s)\rangle=-\langle s_i \rangle +\langle \tanh(\beta \sum_{j\in \mathcal N(i)} J_{ij} s_j + \beta h_i s_i)\rangle
\end{equation}
If we now replace the Hamiltonian for the spin ice, we note that can write coarse grained differential equations by replacing $S_j^A=\langle s_j\rangle$, and defining $m_A(t)=\langle\frac{1}{N} \sum_{j\in A} s_j(t) \rangle$. We obtain as a result the following differential equations:
\begin{equation}
    \dot{\mathbf{m}} =-\mathbf{m}+ \tanh (\beta  \mathbf{Q}\mathbf{m}+\beta \mathbf{h}),
\end{equation}
which generalizes the mean field equations we obtained for the static case to the dynamical regime. For a larger set of mean field parameter we can use also these equations to search for solutions of the static mean field equations.

It is known that for the case of synchronous Glauber dynamics the asymptotic state distribution is described by 
\begin{equation}
    P_{eq}(s)=\mathcal N e^{-\beta \tilde H(s)},
\end{equation}
where $\tilde H=\frac{1}{\beta} \sum_{A} \log 2\cosh(\beta \Delta_a)$ is Peretto's pseudo-Hamiltonian \cite{Peretto} and $\mathcal N$ a normalization constant, with $\Delta_a=J\sum_{B} A_{ab} m_b $.\footnote{The change of sign compared to Peretto is because we are considering an anti-ferromagnetic system.}
For large values of $\beta$ (i.e. low temperature), we have
\begin{equation}
    \tilde H \underbrace{\approx}_{\beta\gg 1}  J \sum_a | \sum_b A_{ab} m_b|
\end{equation}
which is different from the original Hamiltonian, in particular it is linear in the spins, but depends on the absolute value for an arbitrary coupling matrix $J_{ij}$.
Solving for the ground state of the system above can be done via a linear program.

For the case of spin ices this is sufficient to recover the right ground state, as the ground state of a spin ice system is \textit{vertex}-frustrated. We saw at the beginning of the paper that we can write
\begin{equation}
    H=J \sum_v (\sum_{\beta =1} B_{v,\beta }s_\beta)^2.
\end{equation}

It is not surprising then to see
that the ground state (not the excitations) of the system above is equivalent to the one of
\begin{eqnarray}
    \tilde H&=&J  \sum_\alpha  |\sum_v B_{\alpha,v }\sum_{\beta =1} B_{v,\beta }s_\beta | \nonumber \\
    &\leq&   \sum_v  \sum_\alpha J |B_{\alpha v} |\sum_{\beta =1} B_{v,\beta }s_\beta | \nonumber \\
    &\leq&   \sum_v   \tilde J_v |\sum_{\beta =1} B_{v,\beta }s_\beta | \nonumber \\
    &\leq& \tilde J_{max} \sum_v  |\sum_{\beta =1} B_{v,\beta }s_\beta | 
\end{eqnarray}
which is the pseudo-Hamiltonian for the spin ice model from the mean field dynamics. In this sense, as long as we only care about the ground state of the model, the mean field dynamics should provide the right asymptotic ground states; this statement depends on the fact that one can have ground states in which $\sum_{\beta} B_{v\beta} s_\beta=0, \forall v$, which are exactly the ice states. This fact is shown in the next section numerically.

\section{Mean field dynamics: single spin}
The mean field dynamics of an Ising system driven by Glauber dynamics reduce to
\begin{equation}
    \dot{m}(t) = -m(t) + \tanh [\beta Q(m(t) + \beta h(t))]
\end{equation}
for a single spin. We can analyze this simplified system in a variety of parameter regimes to qualitatively understand the limits of the systems responsiveness to external field and the "quality" of the system. The quality is an appropriate measure of the similarity of outputs given similar inputs (kernel quality) minus the similarity of outputs with dissimilar inputs (generalization capability). A natural measure of this emerges from a high temperature analysis while the slow dynamics have simple enough behavior that one may comment on quality by inspection.

\subsection{Slow Dynamics}

When considering this system as a reservoir, it is pertinent to analyze the response of $m(t)$ to $h(t)$. We begin in the case of slow dynamics where $m(t)$ is close to a fixed point and $h(t)$ evolves sufficiently gradually. In this case, the derivative of the mean moment approaches zero, yielding a self consistency equation for the magnetic moment:
\begin{equation}
     m = \tanh [\beta Qm + \beta h].
\end{equation}
This takes the form of a static mean field Ising system. First, let us consider the ferromagnetic case ($Q > 0$). At zero field there is a paramagnetic to ferromagnetic transition when the slope of the hyperbolic tangent evaluated at zero grows equal to one, resulting in a pitchfork bifurcation when $\beta Q = 1$. A single stable solution, $m = 0$, transitions to an unstable $m = 0$ and two stable nonzero solutions resulting from the new intersection of the hyperbolic tangent with the line. The nonzero solutions for $m$ are stable and grow in value with decreasing $\beta$, corresponding to the system settling into a two-fold degenerate ground state. When $h$ is small but not zero, the values of the solutions are shifted in the direction of the field. With a large enough $h$, the unstable and smaller stable points converge to a single unstable fixed point in a saddle node bifurcation. The physical interpretation of this convergence to a single moment in the direction of the field is that the coercivity of the system has been overcome and a hysteritic moment opposing the external field must flip.

\begin{figure}[ht]
    \includegraphics[width=1.0\linewidth]{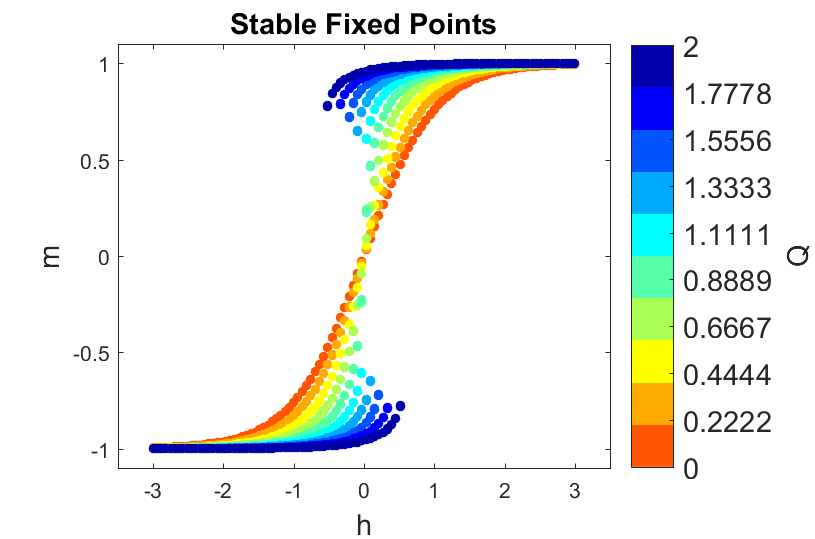}
	\centering
	\caption{Stable fixed points plotted at different temperatures with $Q = 1$. The paramagnetic to ferromagnetic transition is apparent at $k_BT = 1$ as the solutions split into two curves. Magnetic susceptibility and coercivity increases as the temperature is further cooled.
	}
	\label{img:crlt}
\end{figure}

One can infer that limiting this system to slow dynamics constrains it to act like a simple hystersis function. Any applied field can be mapped to an output field if the system's previous state is known. This directly defies the echo state property and constrains the parameter space of the Ising systems used in reservoir computing. If the temperature is too low with slow dynamics, a nonzero coercivity makes for ambiguous computation. This may otherwise be described in the language of the quality of the system. When the coercivity is non-zero, the kernel quality decays as similar inputs may produce different outputs due to the influence of the system's history. Further, whenever the system is close to saturation, the generalization capability falls as different fields in this range all produce a similarly high mean moment. Generally speaking, a slow reservoir computes well when coercivity is low (intermediate temperature) and the system is not near saturation (low field).

The antiferromagnetic case ($Q < 0$) behaves almost precisely like the paramagnetic regime of the ferromagnetic case. When the temperature is further reduced, the shape of the hyperbolic tangent function approaches a step function, leading to a sharper and sharper response of the magnetic moment. This lack of hystersis makes it a more reliable component of a reservoir.

\subsection{High Temperature Limit}
When the temperature is sufficiently high, i.e. where $Q << 1$ and $Qh(t) << 1$, we may linearize the equations of motion and attain exact solutions. Approximating to first order,
\begin{equation}
    \dot{m}(t) = -\lambda m(t) + h(t).
\end{equation}
where the characteristic timescale $\lambda = 1-Q$. This equation equivalently describes the dynamics of a simple RC circuit. An ansatz to this equation ansatz may be written as a homogenous and inhomogenous solution: $m(t) = m_h(t) + m_i(t)$. The homogenous solution, $m_h(t) = A\exp(-\lambda t)$, decays only when $\lambda > 0$. An exponential growth of this term would be unphysical and suppressed by higher order terms in the expansion. The analysis within this section is therefore only valid when $Q < 1$, consistent with the previous assumptions.

The particular solution depends on the timeseries $h(t)$. Assuming this function is periodic with frequency $\omega_h$, we write the Fourier series decomposition:
\begin{equation}
    h(t) = \sum_n C_n \exp(in\omega_h t).
\end{equation}
Matching coefficients for the inhomogenous solution, we find
\begin{equation}
    m_i(t) = \sum_n \frac{1}{in\omega + \lambda} C_n\exp(in\omega t).
\end{equation}
The coefficients are reduced by a factor proportional to the number of the harmonic, indicating the low pass filtering property which becomes most prominent when $n\omega$ and $\lambda$ are of the same order, implying that $\lambda$ is the natural cutoff frequency of the system.

To understand the quality of this system, we calculate how the asympotic and therefore imhomogenous solutions with different driving fields correlate over time. Defining a correlator $\langle m(t) m^{'*}(t) \rangle$ where $m^{'}(t)$ solves 
$\dot{m}^{'}(t) = -\lambda m^{'}(t) + h^{'}(t)$ and $h^{'}(t) = \sum_n C_n^{'} \exp(in\omega_h t)$. $\omega_h$ is the same as the other signal because most reservoir computers fix input lengths and any harmonics that aren't integer multiples of each other will have vanishing overlaps. The average is conducted forward in time:
\begin{equation}
    \langle m(t) m^{'*}(t) \rangle = \lim_{T \to \infty}\frac 1T \int_0^T m(t) m^{'*}(t) dt
\end{equation}
Choosing a starting time arbitrarily late enough such that the homogenous solution approaches zero, only the inhomogenous solution is nonzero.
\begin{equation}
    \langle m(t) m^{'*}(t) \rangle = \lim_{T \to \infty}\frac 1T \sum_{n n^{'}} \int_0^T  \frac{C_nC_{n'}^{*}}{n^2\omega^2 + \lambda^2} \exp(i(n-n^{'})\omega t) dt
\end{equation}
Over a single period, the integrand will always go to zero unless the argument of the exponential goes to zero. This leaves only the terms where $n = n^{'}$ and the overlap simplifies to:
\begin{equation}
    \langle m(t) m^{'*}(t) \rangle = \sum_{n}  \frac{C_nC_{n}^{*}}{n^2\omega^2 + \lambda^2}
\end{equation}
It follows then that at high frequencies this model gives a quality dependency of the type $1/\omega^2$.
\section{Vertices and Vortices}

\subsection{Vertices}
The famed degeneracy of artificial spin ice is often analzyed through the lens of energetics. That is, extensively numerous states of the system possess the same energy. The states are categorized at vertices, natural building blocks of the system. We analyze an isolated vertex from the square ice system to observe how ice properties may be achieved dynamically and what the transitions between states imply. To model this system, we label all four $m_i = 1$ states as pointing up and to the right, indexed counterclockwise from the rightmost magnet. The external field is applied equally to all magnets along the same direction. The interactions between magnets are parameterized in terms of $J_1$, the interaction between spins oriented perpendicular to one another which is typically stronger in planar square ice, and $J_2$, the typically weaker interaction of collinear macrospins. The interaction matrix is thus
\begin{equation}
    \mathbf{Q} = \begin{pmatrix}
q & -J_1 & J_2 & J_1\\
-J_1 & q & J_1 & J_2\\
J_2 & J_1 & q & -J_1\\
J_1 & J_2 & -J_1 & q\\
\end{pmatrix}
\end{equation}
and the field vector is
\begin{equation}
    \mathbf{h}(t) = h(t)\begin{pmatrix}
    1\\
    1\\
    1\\
    1
    \end{pmatrix}.
\end{equation}

\subsection{Low magnetization fixed point}
Considering the case a small argument to the $\tanh$ function and a static field $h(t) = h$, one finds a fixed point at $m_i = \frac{h}{q+J_2-1/\beta}$ for all $i$. This intuitively implies that the fixed point is slightly offset from 0 by the external field, shifting the entirety of the phase space and therefore allowing the system to switch between basins of attraction. There is a critical temperature where $1/\beta = q +J_2$ that maximizes the sensitivity to external field, implying that this is an ideal temperature for reservoir computing as only small amounts of driving are required to alter the entire system's state. This will be explored further in the numerical analysis of the phase space.

The eigenvectors of the Jacobian about the saddle node are all associated with real eigenvalues that change sign depending on the system parameters. Positive eigenvalue directions are evolved towards and typically manifest as the system approaching a high $m_i$ state of a energetically favorable vertex type. Specifically, the eigenvectors and their corresponding eigenvalues are the following:
\begin{equation}
    \mathbf{v}_1 = \begin{pmatrix}
    1\\
    -1\\
    -1\\
    1
    \end{pmatrix}, 
    \mathbf{v}_2 = \begin{pmatrix}
    1\\
    0\\
    1\\
    0
    \end{pmatrix}, 
    \mathbf{v}_3 = \begin{pmatrix}
    0\\
    1\\
    0\\
    1
    \end{pmatrix}, 
    \mathbf{v}_4 = \begin{pmatrix}
    -1\\
    -1\\
    1\\
    1
    \end{pmatrix}.
\end{equation}
The corresponding eigenvalues are
$$\lambda_1 = \beta(q + 2J_1 -J_2) - 1,$$
$$\lambda_2 = \lambda_3 = \beta(q + J_2) - 1,$$  
\begin{equation}
    \lambda_4 = \beta(q - 2J_1 -J_2) - 1.
\end{equation}
The first eigenvector is precisely the direction of a Type I vertex, the second and third comprise collinear spin ordering that together lead to Type II states, and the fourth direction is a type IV vertex. In all cases, beta may be reduced low enough to make the eigenvalues negative, effectively raising the temperature high enough to "melt" the individual spins into paramagnets and make the fixed point stable. On the other extreme, the self energy $q$ dominates the interactions and the fixed point is unstable. When parameters are similar to experiments, that is, $\beta q >> 1$, and the rest of the parameters are on the same order of magnitude, the relative strength of $J_1$ and $J_2$ makes the Type I and Type II associated eigenvalues respectively more positive, leading towards a faster and more dominant evolution of the system towards those vertex types.

\subsection{Basins of attraction}
To characterize the phase space, we initialized the system across evenly spaced points on the physically relevant interval [-1,1]. We then evolved the system over time until the system converged and recorded the closest Ising values of $m_i$ and average net moment, $\langle m_i^2\rangle$. We then categorize the Ising states into vertex types and record the average fraction of vertex types as parameters are varied \ref{img:MFDTV}).

The temperatures between $10^{-1}$ and $10^{1}$ represent the expected behavior for a spin ice. The Type I and II vertices occur proportional to the 1:2 ratio of microstates they comprise while all other states are near zero. Below $10^{-1}$, a new type III state emerges due to the ability to partially magnetize individual spins. Finding a low energy configuration through reducing the moments' magnitudes is not found in nanoscale experiments, potentially due to the inability to appropriately cool the systems, but does occur in macroscopic dipole spin ice.

\begin{figure}[ht]
    \includegraphics[width=1.0\linewidth]{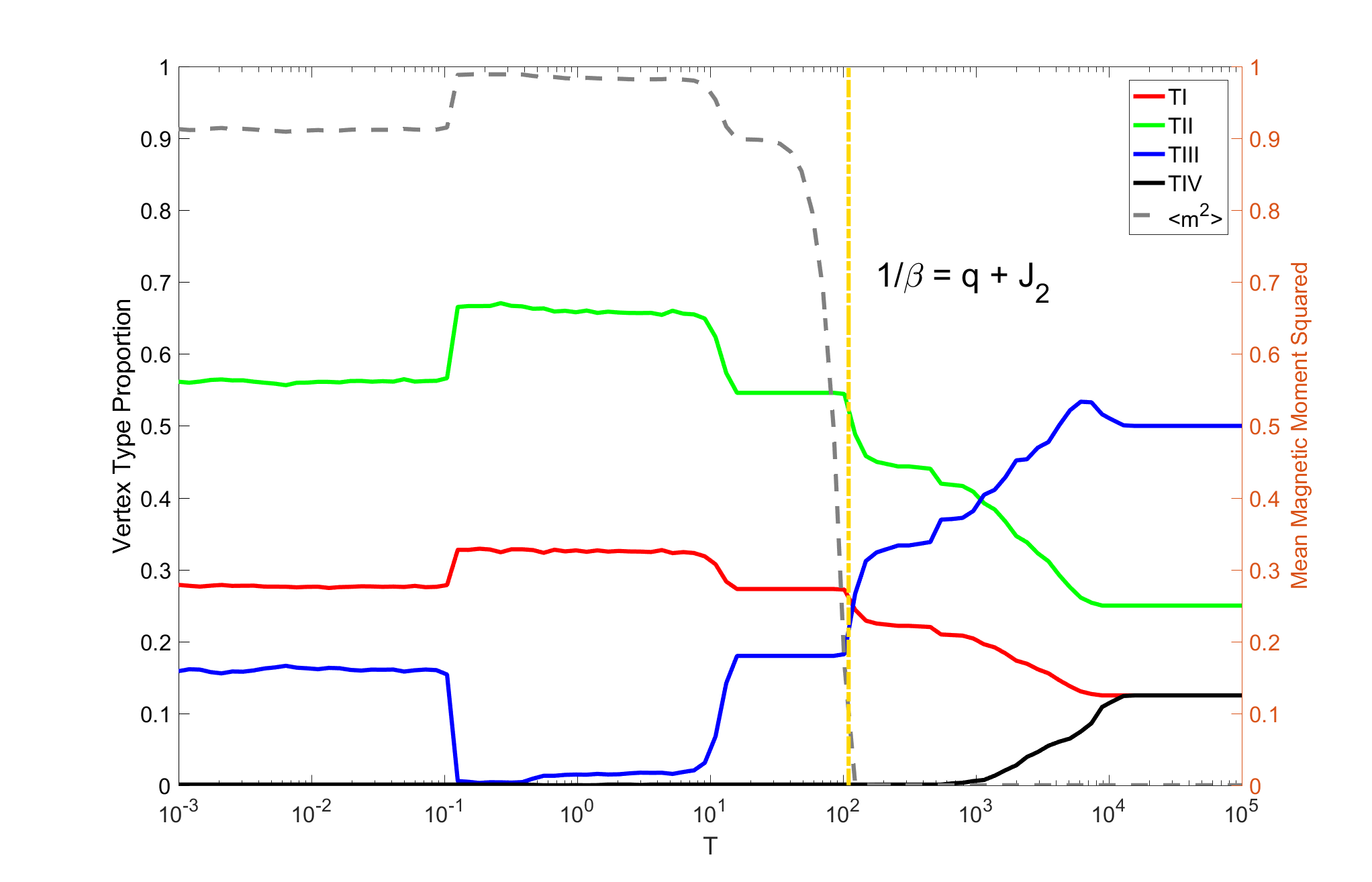}
	\centering
	\caption{Fraction of the system evolving towards vertex types as a function of temperature for $J_1 = 100$, $J_2 = 100$, $q = 10$, and $h = 0$. Both the fraction and average magnetization parameter, $\langle m_i^2\rangle$, are plotted on the same scale. The critical temperature of maximum fluctuation is plotted at the goldenrod line. 
	}
	\label{img:MFDTV}
\end{figure}

\begin{figure*}[ht]
    \includegraphics[width=1\linewidth]{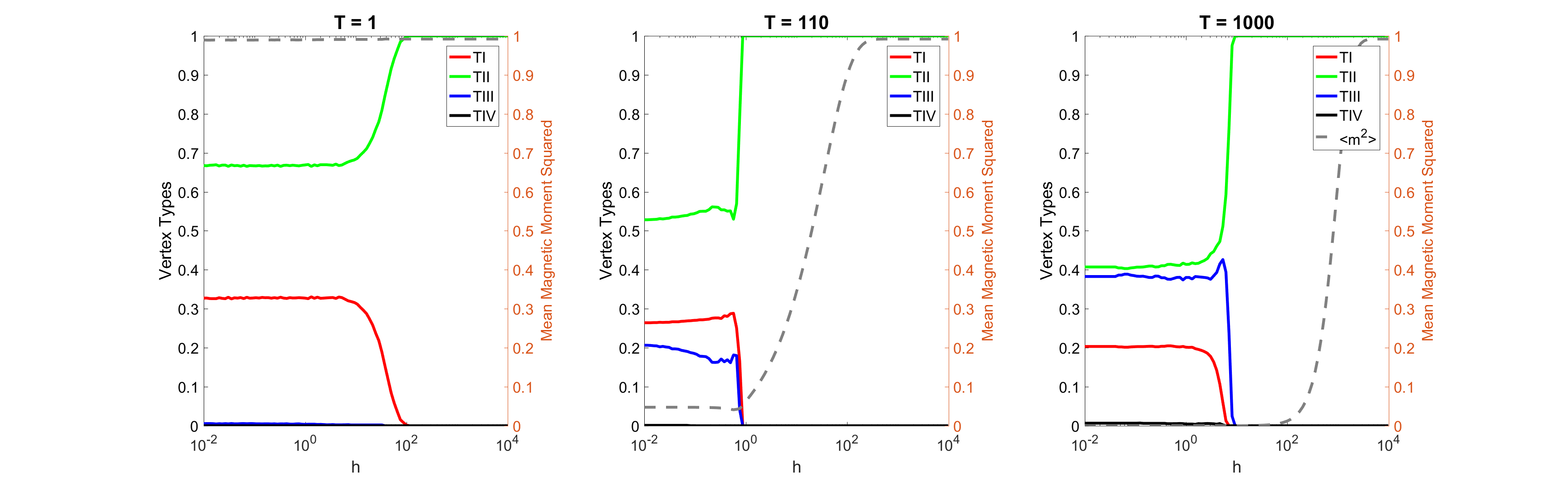}
	\centering
	\caption{Vertex types at a range of field strengths for $J_1 = 100$, $J_2 = 100$, $q = 10$, and three separate temperatures.
	}
	\label{img:MFDTV}
\end{figure*}

\begin{figure}[ht]
    \includegraphics[width=1\linewidth]{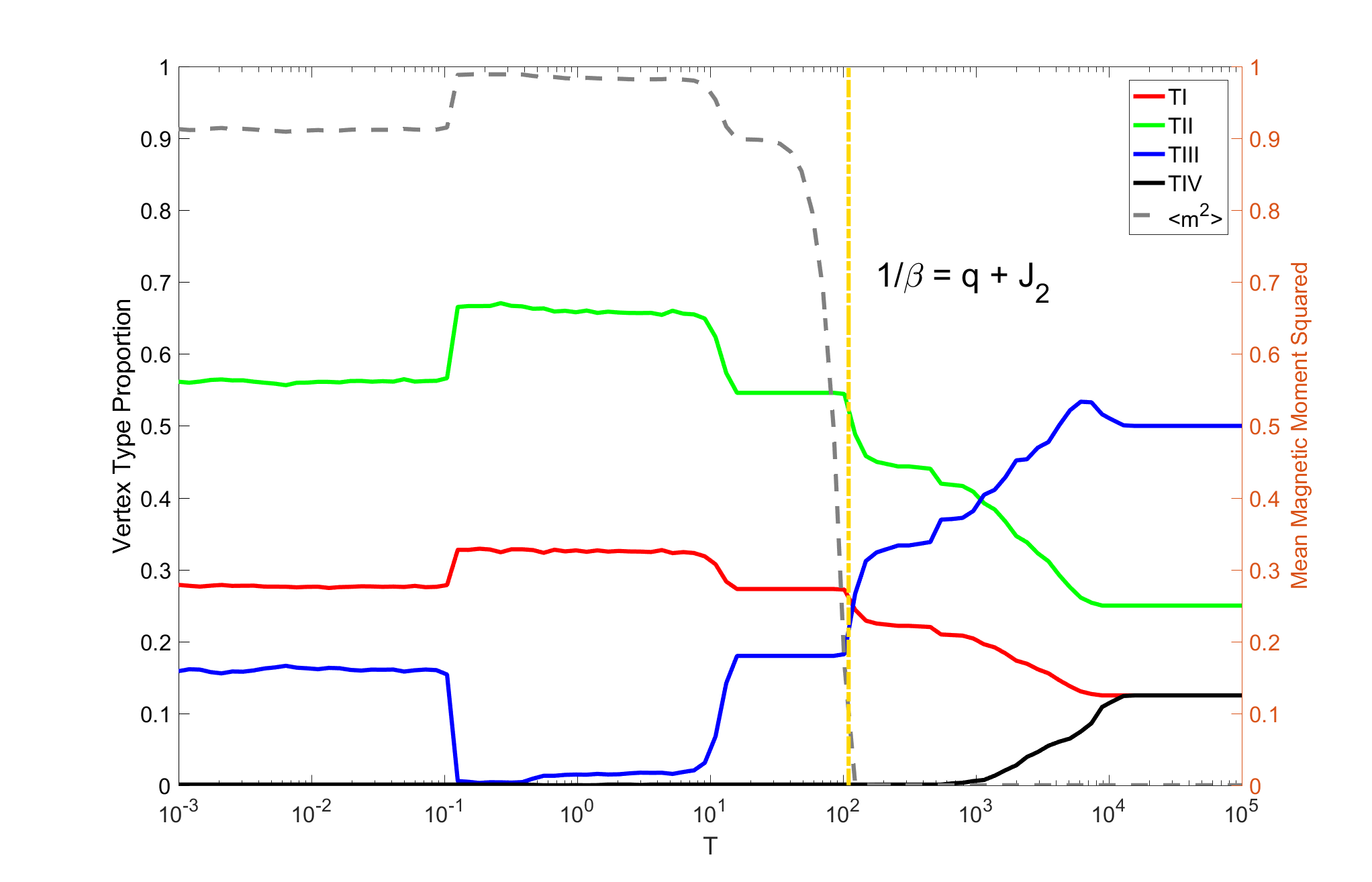}
	\centering
	\caption{Fraction of the system evolving towards vertex types as a function of temperature for $J_1 = 100$, $J_2 = 100$, $q = 10$, and $h = 0$. Both the fraction and average magnetization parameter, $\langle m_i^2\rangle$, are plotted on the same scale. The critical temperature of maximum fluctuation is plotted at the goldenrod line. 
	}
	\label{img:MFDTV}
\end{figure}

\subsection{Vortices}

\begin{figure}
    \centering
    \includegraphics[scale=0.6]{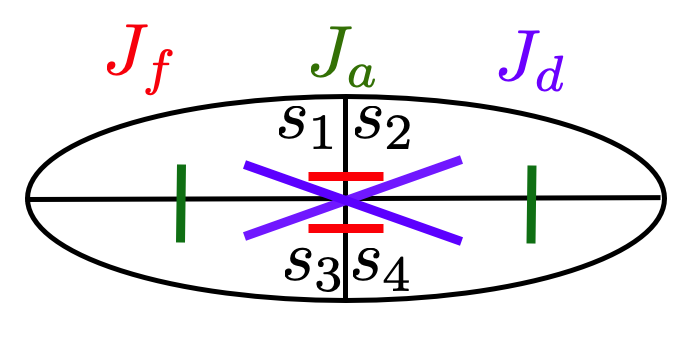}
    \caption{Interactions for a single island: we have diagonal, horizontal and vertical interactions.}
    \label{fig:intisland}
\end{figure}

Recent experiments have shown the potential for real spin islands to develop vortex states: spin textures no longer dominated by a single direction but rather spins that wrap around the center of the nanomagnet. A simple model to capture the emergence of a vortex in this mean field format is one in which the island is divided into four quadrants, each with an individual mean field spin. By symmetry, there are three interaction strengths between these quadrants: $J_f$ between collinear spins with ferromagnetic dipolar interactions, $J_a$ between parallel, non-collinear spins with antiferromagnetic dipolar interactions, and $J_d$ between spins across the diagonal from one another. In the convention that all spins are positive when pointing to the right, the interaction matrix is
\begin{equation}
    \mathbf{Q} = \begin{pmatrix}
q & J_f & J_d & J_a\\
J_f & q & J_a & J_d\\
J_d & J_a & q & J_f\\
J_a & J_d & J_f & q\\
\end{pmatrix}
\end{equation}
Applying the same fixed point analysis as before, we find that all $m_i = \frac{h}{q+J_d+J_f-J_a}$ and the following eigenvectors of the Jacobian:
\begin{eqnarray}
    \mathbf{v}_{mag} &=& \frac{1}{2}\begin{pmatrix}
    1\\
    1\\
    1\\
    1
    \end{pmatrix}, 
    \mathbf{v}_{1vort} = \frac{1}{2}\begin{pmatrix}
    -1\\
    -1\\
    1\\
    1
    \end{pmatrix}, \\
    \mathbf{v}_{nomag} &=& \frac{1}{2}\begin{pmatrix}
    -1\\
    1\\
    -1\\
    1
    \end{pmatrix}, 
    \mathbf{v}_{2vort} = \frac{1}{2}\begin{pmatrix}
    1\\
    -1\\
    -1\\
    1
    \end{pmatrix},
\end{eqnarray}
where the factor $\frac{1}{2}$ is for normalization.
The corresponding eigenvalues are
\begin{eqnarray}
\lambda_{mag}&=&\beta(q + J_a + J_f + J_d) - 1,\nonumber \\
\lambda_{1vort} &=& \beta(q - J_a + J_f - J_d) - 1,\nonumber \\
\lambda_{nomag} &=& \beta(q - J_a - J_f + J_d) - 1,\nonumber \\
    \lambda_{2vort} &=& \beta(q + J_a - J_f - J_d) - 1.
\end{eqnarray}
\subsubsection{Statics}
We note that the Hamiltonian can be written in the form
\begin{eqnarray}
    H_{island}=\frac{1}{2} \vec s^t\ \mathbf{Q} \vec s
\end{eqnarray}
with $\vec s=(s_1,s_2,s_3,s_4)$, and that thus the eigenvectors are the same. Let us call $\rho_k$ the corresponding eigenvalues. We have $\lambda_k=\beta \rho_k-1$. The semi-magnetized states can be written then as superpositions of the magnetized, unstructured unmagnetized, and the 1 and 2 vortices solutions.

This also implies that, if we call $c_k=\vec s\cdot \vec v_k$, we can rewrite

\begin{eqnarray}
    H_{island}&=&\frac{1}{2}(\rho_{mag} c_{mag}^2+\rho_{nomag} c_{nomag}^2\\
    &+&\rho_{1vort} c_{1vort}^2+\rho_{2vort} c_{2vort}^2)
\end{eqnarray}
Minimizing the Hamiltonian is thus associated to the minimum eigenvalue as a function of $J_a,J_f,J_d$. Since $q$ is a shift, we can set it to zero in this case. Moreover, we have the freedom of setting $J_{f}=-1$, thus reducing the phase diagram to $J_{d}$ and $J_{a}$ in units of $J_{f}$. This is equivalent to the minimum eigenvalue problem for the matrix
\begin{equation}
    \mathbf{\tilde Q} = \begin{pmatrix}
0 & -1 & -\tilde J_d & -\tilde J_a\\
-1 & 0 & -\tilde J_a & -\tilde J_d\\
-\tilde J_d & -\tilde  J_a & 0 & -1\\
- \tilde  J_a & -\tilde  J_d & -1 & 0\\
\end{pmatrix}
\end{equation}
with eigenvalues
\begin{eqnarray}
    \tilde \rho_{mag}&=&-\tilde J_a-\tilde J_d-1\\
    \tilde \rho_{nomag}&=&\tilde J_a-\tilde J_d+1 \\
        \tilde \rho_{1vort}&=&\tilde J_a+\tilde J_d-1\\
    \tilde \rho_{2vort}&=&-\tilde J_a+\tilde J_d+1
\end{eqnarray}

\subsubsection{Dynamics}
For the dynamics, however, $q$ becomes important as it affects the role that that temperature has.
We can largely draw the same conclusions with respect to the $q$ dependence: larger values of $q$ drive all eigenvalues towards positivity which in turn makes all modes unstable, leading to an evolution towards ergodic states. The 4th state is the first to become stable and therefore cease to exist in an ensemble of systems. This state is effectively the island splitting into two macrospins aligned energetically unfavorably with one another. The remaining three states (1 being a macrospin, 2 being a vortex, and 3 being two vortices) emerge when two interaction strengths combined are larger than the third. This allows for coexistence of these three states across a wide variety of interaction ranges while the precise size of their basins of attraction are tunable via the balance of interaction strengths (Fig. 5). 

One may additionally observe a point of maximum sensitivity of the fixed point to external field: $q = J_a - J_d - J_f$. This is especially relevant to the exploitation of vortex formation in FMR facilitated reservoir computing. The prediction of double vortices with appreciable populations should allow another form of system output which in term increases the amount of processing the system is capable of.

\subsection{Square and Hexagonal Ices: Glauber vs. Mean field dynamics}
Let us now check that this model reproduces also the ground states of the Hexagonal Artificial Spin Ice (HeX), and the Artificial Square Ice (ASI).

\begin{figure*}
\includegraphics[scale=0.2]{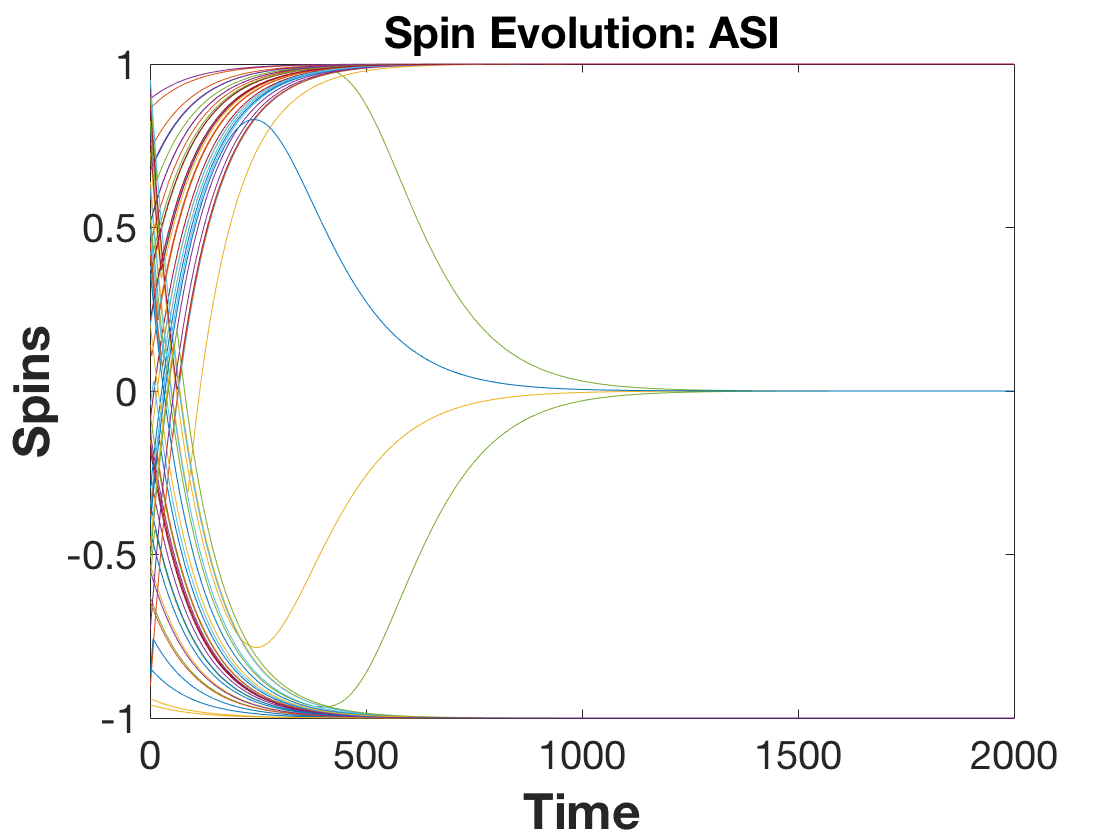}\includegraphics[scale=0.2]{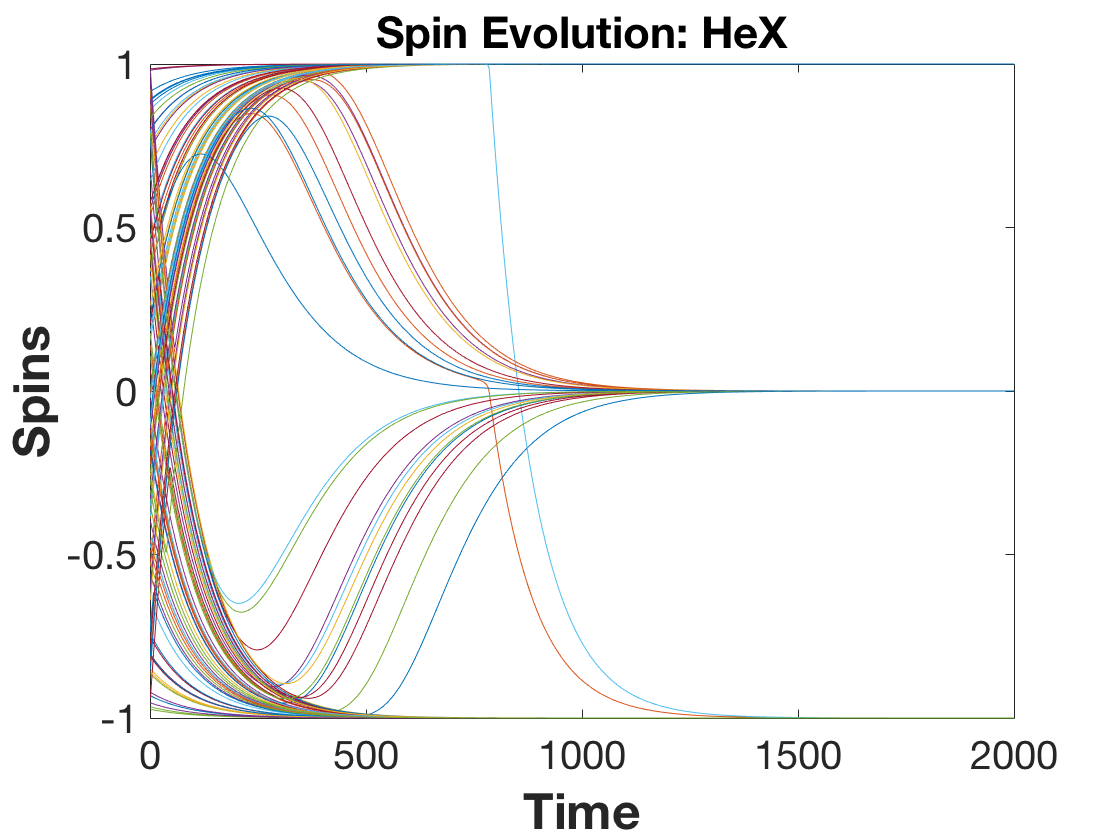}\\
\includegraphics[scale=0.2]{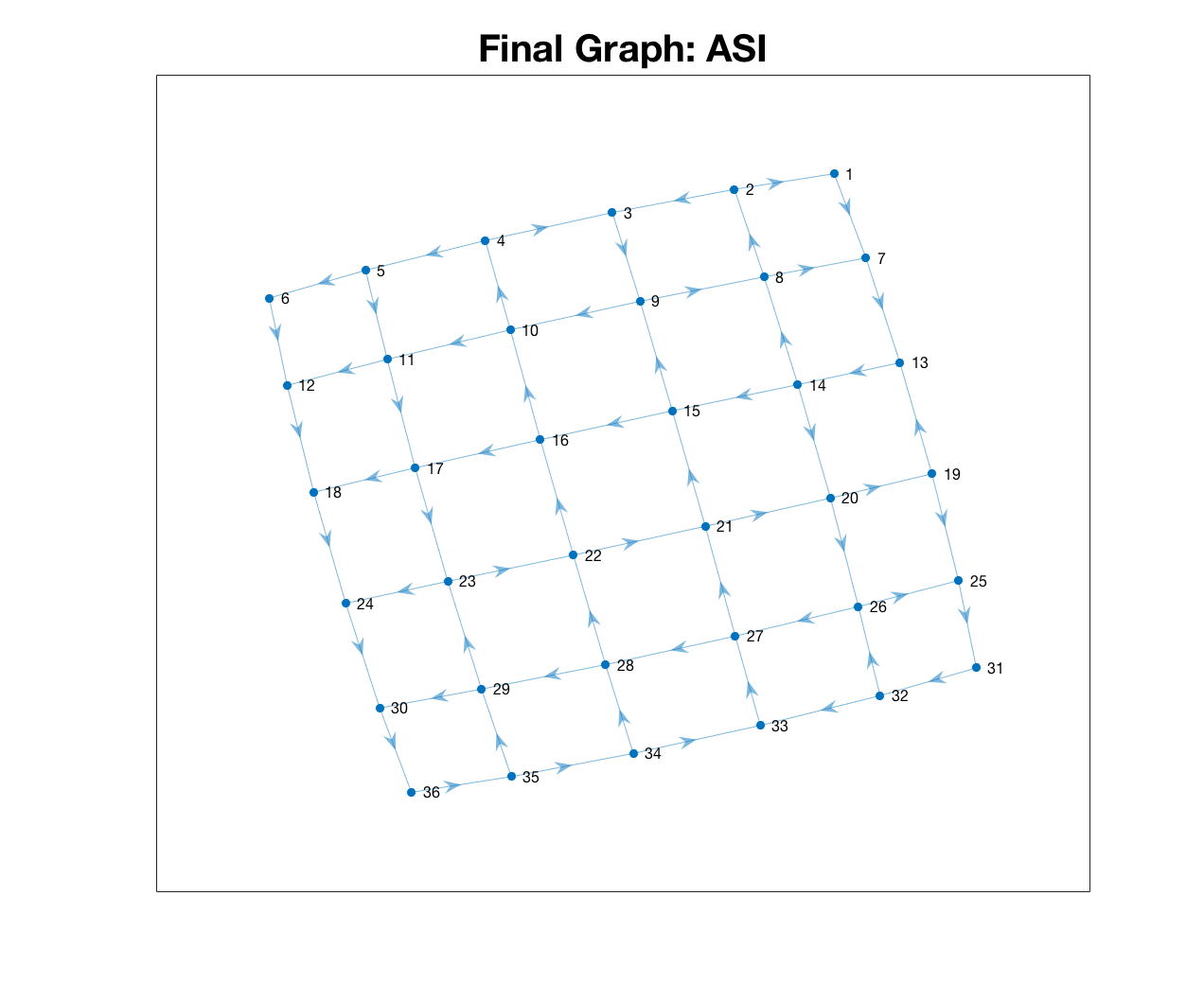}\includegraphics[scale=0.2]{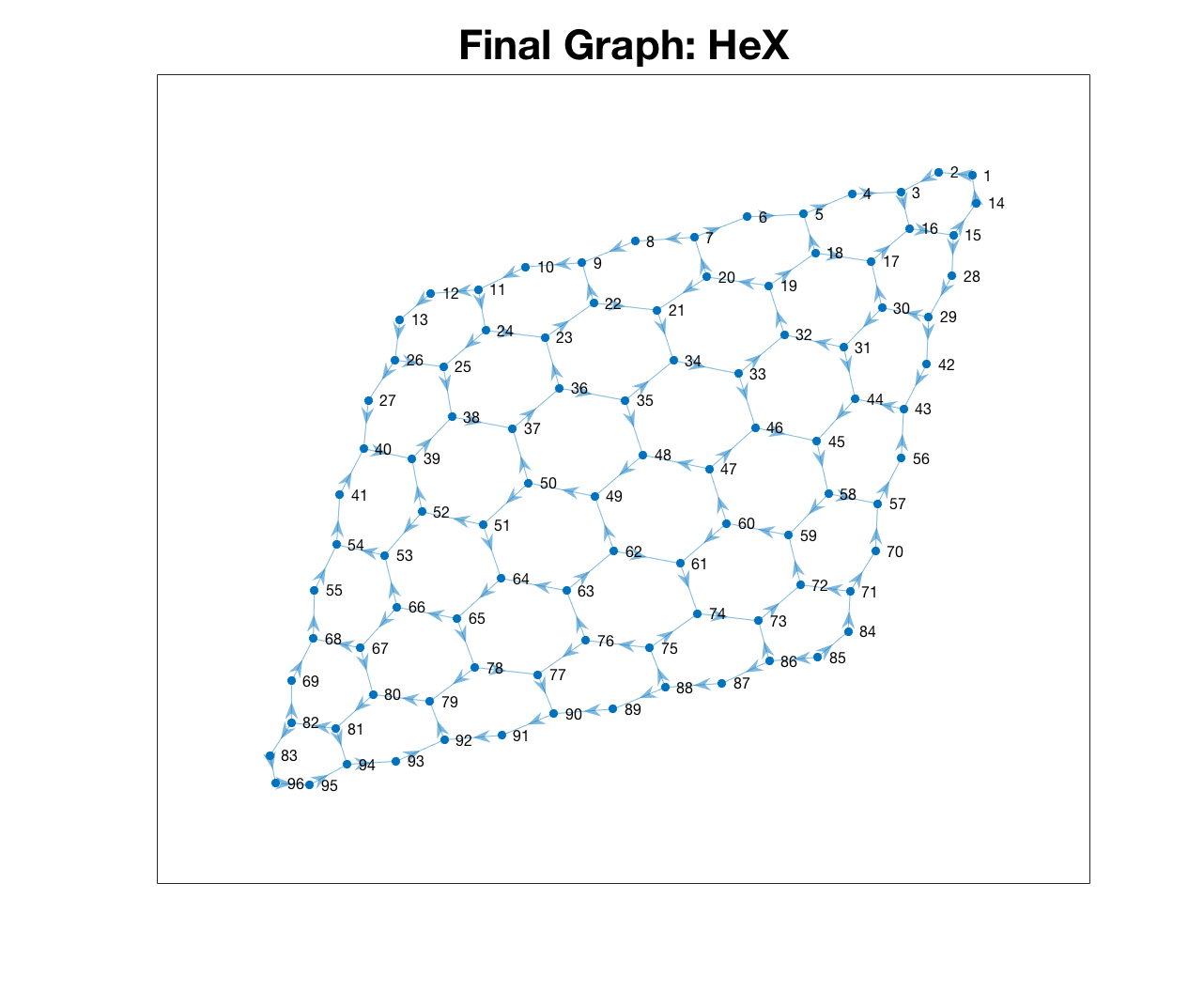}\\
\caption{Relaxation of each spin of the Artificial Square Ice (ASI) and the Hexagonal Spin Ice (HeX) in the mean field approximation below the ordering temperatures $T_o$, and the corresponding final states. It can be seen that the system converges to the known correct ground state for the two systems.}
\label{fig:ordering}
\end{figure*}

The first thing to notice is that, given the eigenvalues of the effective Ising interaction matrix $J_{ij}$, the ordering occurs for temperatures below the maximum eigenvalues $T_o=\Lambda_{max}$. Above the temperature $T_o$, the system is in a paramagnetic phase and the spins converge exponentially to $m=0$. At temperatures below $T_o$, we observe an interesting phenomenon, which we describe in Fig. \ref{fig:ordering} for \textit{HeX} and \textit{ASI}. We observe that as time evolves, below the ordering temperatures the Ising models
some of the spins will converge towards values $+1$ and $-1$. Some of the spins instead converge towards the value zero. The asymptotic states depends on the initial condition. These spins are those that can be freely flipped without leaving the ice manifold, and we can thus assign to them random positive and negative spin values. Given this arrangement, in Fig.  \ref{fig:ordering} we show the final configuration both for ASI and HeX, which can be promptly checked to be in the ice manifold. 

At the temperature $T_o$, the mean field equations still converge to $m=0$, but the regime is not exponential anymore, but a power law which can be extrapolate to be roughly $\sqrt{t^{-1}}$, which is the mean field dynamical exponent. A well known example of possible application of artificial square ice is the possibility of having monopole like charges in  spin ices \textit{without} a string tension. Such degeneracy can be obtained via mediated spin interactions \cite{caravellimed} or staggered island heights \cite{farhan2019emergent}.

\begin{figure}
    \centering
    \includegraphics[scale=0.2]{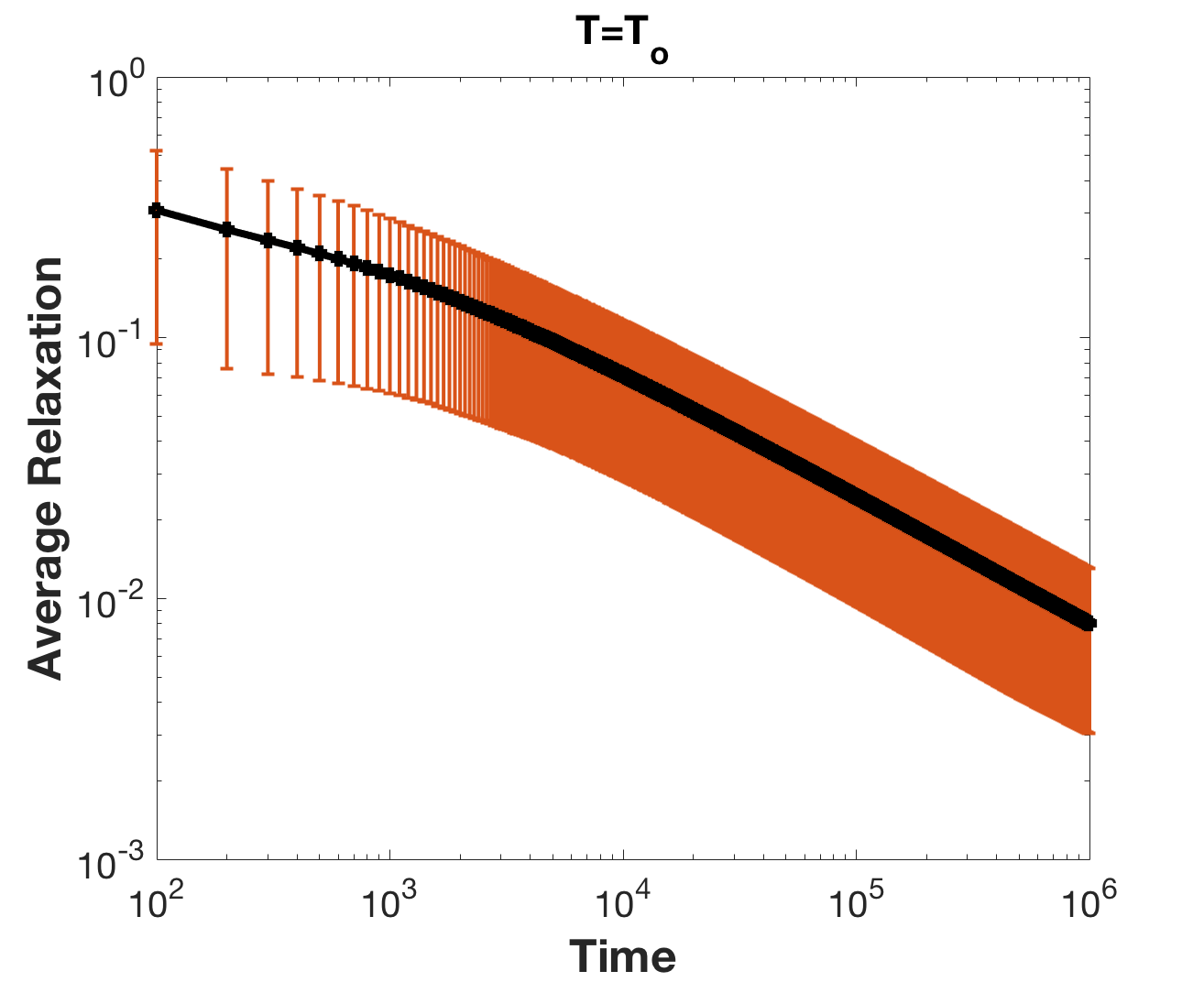}
    \caption{Mean field magnetization of each spin of an ASI  at $T=T_o$. We observe that the average spin $m_{av}(t)=\frac{1}{N} \sum_i |m_i(t)|$ relaxes as a power law $t^\alpha$ , with exponent roughly $\alpha \approx 0.5$ consistent with the mean field approximation.}
    \label{fig:my_label}
\end{figure}

\subsection{Embedding Logic gates}
Let us also consider another application of this model, e.g. logic gates embedded in magnetic nanoislands. As discussed in \cite{caravellinislg}, the asymptotic states are not necessarily in the spin ice manifold, but we find interesting to see whether the mean field dynamics can be implemented to obtain better results, compared to the discrete Glauber approach, to the ground designed ground state of these models.

The gate functionality can be obtained \cite{caravellinislg} by studying a simple spin Hamiltonian of the form
\begin{equation}
H_3=\sigma_3\left( J(\sigma_1+\sigma_2)+h\right),
\label{eq:gates}
\end{equation}
which we assume that the spins $\sigma_1$ and $\sigma_2$ are inputs, and $\sigma_3$ is our output and free to fluctuate at a certain temperature $T$. If $2|J|>h$ the ground state of the system are those of a logic gates (N)AND and (N)OR, which is why this embedding is successful in reproducing a wide range of logic gates.

The hyerarchy problem
can be solved by modulating  couplings $J$ among islands. For the model without horizontal (input) spin interactions, we choose $\epsilon$ such that $|{J_k}/({2+\epsilon})|<|h_{k }|$. 
As in \cite{caravellinislg}, we used the coupling rescaling between layers given by
\begin{equation}
|J_{k-1}|\leq \frac{|J_{k}|}{(2+\epsilon)}.
\label{eq:coppresc}
\end{equation}

We performed numerical simulations of randomly chosen gates via a Glauber \cite{glauber} spin dynamics with exponential annealing for a system whose couplings scale as in Eq.~(\ref{eq:coppresc}), versus the mean field dynamics results.  The probability of a spin flip in the Glauber model is associated to the  energy change $\Delta E$, and is given by
$p(k)=(1+e^{\Delta E /T_k(t)})^{-1}.$

\begin{figure}
    \centering
    \includegraphics[scale=0.15]{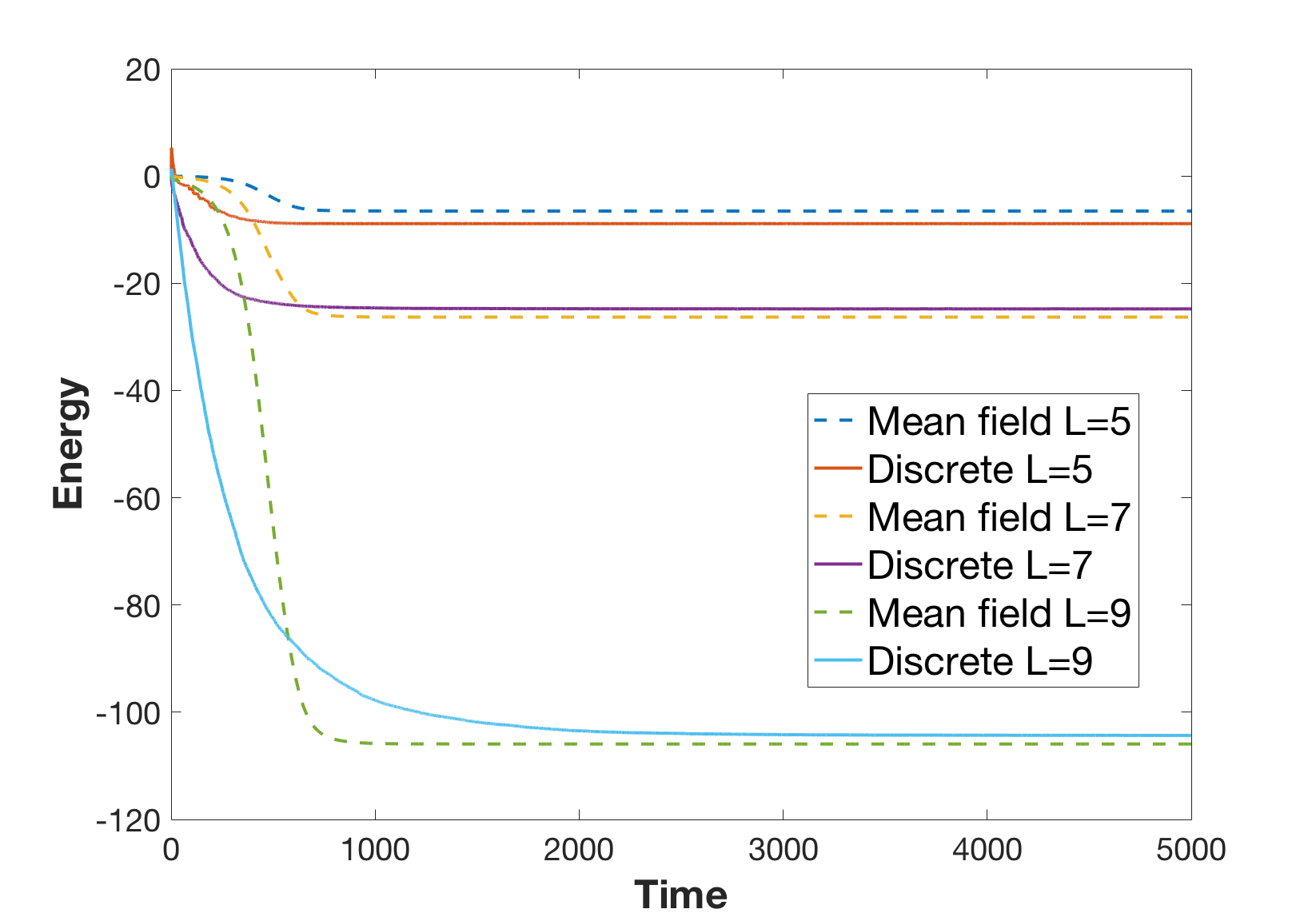}
    \includegraphics[scale=0.15]{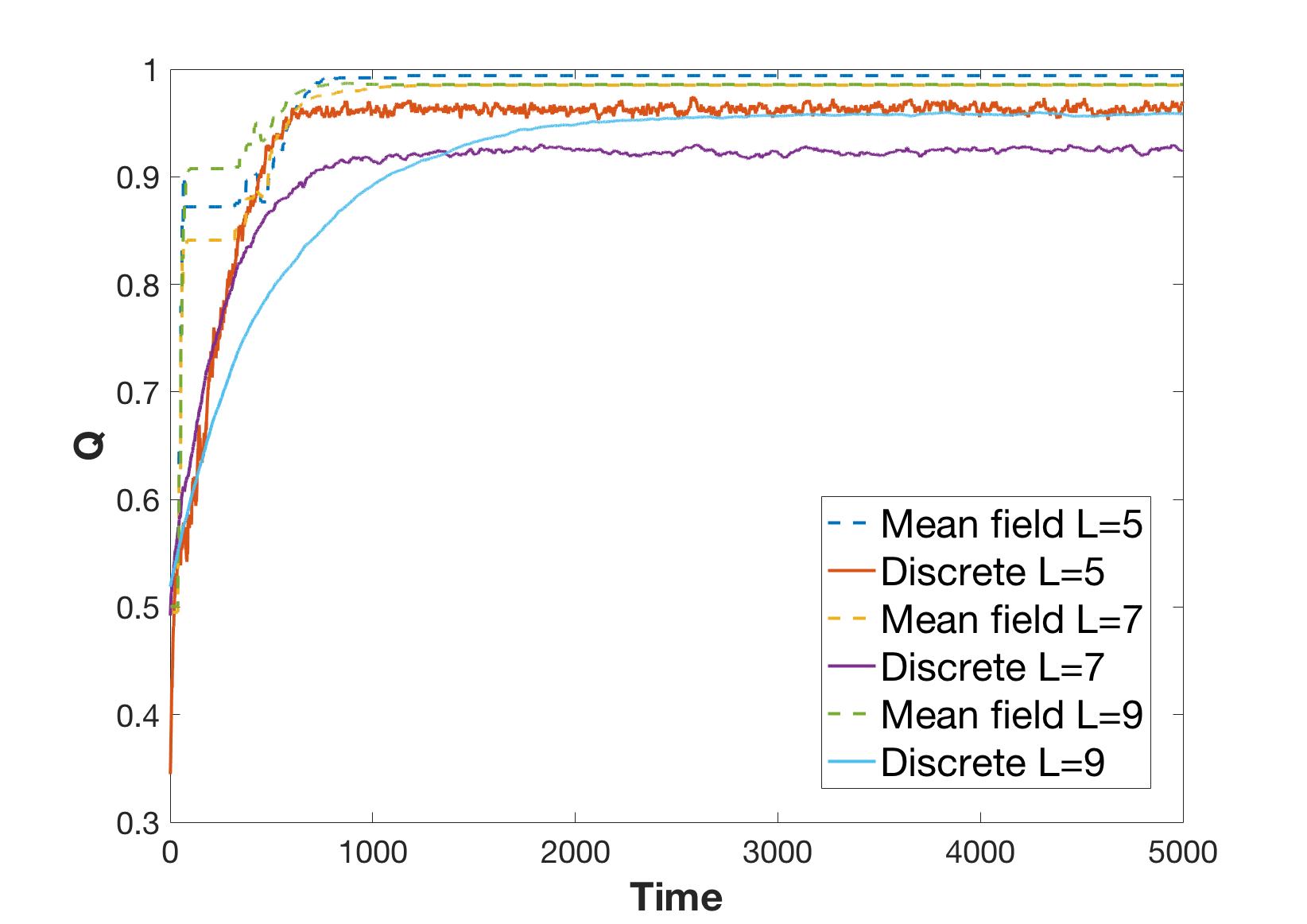}
    \caption{Average corresponding energy and overlap for random gates in the mean field approximation. We oberve that in the mean field approximation, the average overlap is close to $1$ (full functionality of the gates) and higher than the Glauber result for trees of depth $L=5,7,9$. Each line is averaged over 100 simulations for the Glauber dynamics with discrete spins, and $100$ initial conditions for the mean field corresponding equations. The annealing rate is identical for both systems.}
    \label{fig:overlap}
\end{figure}

As in previous papers \cite{caravellinislg}, the control parameter for the fidelity of the circuit we introduce the overlap between the spin system and the equivalent logic circuit, not only for final output, but  {\it for all the intermediate layers}. For each gate, we consider the metric given by the \textit{gate overlap} as the quantity,
\begin{equation}
\mathcal Q=\frac{\left[\vec S\cdot\vec L+(1-\vec S)\cdot(1-\vec L)\right]}{N},
\end{equation} 
where $N$ is the total number of gates. Full functionality corresponds to $\mathcal Q=1$, and completely random system have $Q=0$. In Fig. \ref{fig:overlap} we compare the mean field and Glauber dynamics results, showing that the results are consistent with previous studies.

\end{document}